\documentclass[twocolumn,aps,floatfix,showpacs,pra,superscriptaddress]{revtex4-2}

\usepackage{amsmath}
\usepackage{amssymb}
\usepackage{esint}
\usepackage{graphicx}
\usepackage{bbold}
\usepackage[normalem]{ulem}
\usepackage{physics}
\usepackage{soul}
\usepackage[makeroom]{cancel}
\usepackage{placeins}

\usepackage{setspace}

\usepackage{breakurl}
\usepackage{seqsplit}
\usepackage[hidelinks,colorlinks]{hyperref}
\hypersetup{citecolor=[rgb]{0.25,0.14,0.63}}
\hypersetup{urlcolor=[rgb]{0.25,0.14,0.63}}
\hypersetup{linkcolor=black}

\makeatletter

\usepackage{color}

\begin{document}

\title{The fate of the ``vacuum point'' and of grey solitons in dispersive quantum shock waves in a one-dimensional Bose gas}
\author{S. A. Simmons}
\affiliation{School of Mathematics and Physics, The University of Queensland, Brisbane, Queensland 4072, Australia}
\author{J. C. Pillay}
\affiliation{Quantum Brilliance Pty Ltd, 60 Mills Road, Acton ACT 2601, Australia}
\author{K. V. Kheruntsyan}
\affiliation{School of Mathematics and Physics, The University of Queensland, Brisbane, Queensland 4072, Australia}

\date{\today}

\begin{abstract}
We continue the study of dispersive quantum shock waves in a one-dimensional Bose gas beyond the mean-field approximation. In a recent work by Simmons \textit{et al.} [Phys. Rev. Let. \textbf{125}, 180401 (2020)], the oscillatory shock wave train developing in this system from an initial localized density bump on a uniform background was interpreted as a result of quantum mechanical self-interference, wherein the interference contrast would diminish with the loss of matter-wave phase coherence. Such loss of coherence, relative to the mean-field Gross-Pitaevskii description, occurs due to either quantum or thermal fluctuations, as well as in the strongly interacting regime. In this work, we extend the analysis of dispersive quantum shock waves in this context to other dynamical scenarios. More specifically, the scenarios studied include evolution of a sufficiently high density bump, known to lead to the so-called ``vacuum point''  in the mean-field description, and evolution of an initial density dip, known to shed a train of grey solitons in the same mean-field approximation. We study the fate of these nonlinear wave structures in the presence of quantum and thermal fluctuations, as well as at intermediate and strong interactions, and show that both the vacuum point and grey solitons cease to manifest themselves beyond the mean-field approach. On the other hand, we find that a vacuum point can occur in an ideal (noninteracting) Bose gas evolving from a ground state of a localized dimple potential. Due to the ubiquity of dispersive shock waves in nature, our results should provide useful insights and perspectives for a variety of other physical systems known to display nonlinear wave phenomena.
\end{abstract}
\maketitle

\section{Introduction}

During the last two decades there has been a keen interest, which has peaked again just recently, in the study of the formation and propagation of dispersive quantum shock waves in ultra-cold atomic gases and Bose-Einstein condensates \cite{Dutton2001,Kamchatnov2002,Kamchatnov2004,Damski2004a,Damski2004,Perez-Garcia2004,Simula2005,El2006,Damski2006,Cornell2006,Carusotto2006,Kamchatnov2006,Kamchatnov2007,Hoefer2008,Engels2008,Davis2009,Engels2009,Abanov2012,Crosta2012,Peotta2014,Peotta2014b,Salasnich2016,Hoefer2016,Engels2020,Simmons2020,Dubessy2021,Kamchatnov2021}. This is, in part, due to the high degree of experimental control that we have over these systems, making them an ideal platform to realize scenarios best suited to understand the rich out-of-equilibrium phenomena that arise in interacting quantum many-body systems. Exploring these phenomena in one-dimensional (1D) Bose gases has an added benefit that comes with the integrability of the underlying Lieb-Liniger model and the availability of various theoretical tools that make dynamical many-body simulations computationally tractable.

The focus on dispersive quantum shock waves in ultra-cold Bose gases originates largely through the use of the mean-field Gross-Pitaevskii equation (GPE). Such a mean-field treatment, however, ignores the effects of quantum fluctuations and correlations, despite the fact that the phenomenon under investigation is referred to as a `quantum' shock wave. On the other hand, the 1D GPE is a specific form of the widely applicable non-linear Schr\"odinger (NLS) equation \cite{Karjanto2020} and is closely related to the Korteweg--de Vries (KdV) equation \cite{KdV1895}, both of which are staple foundations used in the hydrodynamic description of wave breaking in dispersive media and soliton formation in nonlinear media \cite{Gurevich1974,Kamchatnov2000,Hoefer2016,Kamchatnov2021}. 
Moreover, in the Thomas-Fermi or weakly-dispersive regime of the GPE, Whitham modulation theory \cite{whitham1974linear} provides an approximate analytic description \cite{Kamchatnov2021} of the formation, shape, and internal structure of dispersive shock waves, allowing for deep analytical insights to be obtained in these matter-wave systems.

In this work, we continue the study of quantum shock wave scenarios in a 1D Bose gas that extend beyond the weakly-dispersive, mean-field GPE regime. More specifically, we study the dissolution of a localized density bump on an otherwise uniform background through dispersive shock waves and extend the analysis of Ref.~\cite{Simmons2020} to the case of a relatively large bump amplitude. At the mean-field GPE level, this scenario is known to lead to the formation of a vacuum point wherein the fluid density becomes zero at some point inside the shock wave train. Here, we go beyond the mean-field GPE description by incorporating the effects of quantum or thermal fluctuations using, respectively, the truncated Wigner (TWA) or classical-field stochastic Gross-Pitaveskii (SPGPE) \cite{Castin2000,Davis2001,Blakie2008} approaches. Additionally, we treat the regimes of intermediate and strong interactions using infinite matrix product states (iMPS). We show that all these beyond-mean-field effects lead to the disappearance of the vacuum point. On the other hand we find that the vacuum point can exist in the limit of a noninteracting (ideal) Bose gas at zero temperature.

In addition to the density bump scenario, we also study the dynamics of the dissolution of a localized density dip. This scenario is known to lead to a train of grey solitons at the mean-field GPE level, however, we show that incorporating the effects of quantum and thermal fluctuations, as well as strong inter-particle interactions, can strongly modify the shape of the solitons or indeed cease their very formation.

In all scenarios, apart from analyzing the dynamics of the mean particle number density $\rho(x,t)$ as in Ref.~\cite{Simmons2020}, we also monitor the evolution of the mean current density $j(x,t)$. The current density arises naturally in the context of the mean-field superfluid hydrodynamic equations [see Eqs. \eqref{eq:QHD-rho_j} and \eqref{eq:QHD-j} below], which are mathematically equivalent to the GPE, and serves as an important and fundamental tool in the field of transport and non-equilibrium physics. In particular, it has been used to characterize the dynamics of ultra-cold Bose gases in scenarios where shock waves are generated by pushing a homogeneous gas against a hard wall boundary \cite{Peotta2014b,Dubessy2021}. In this situation the gas sloshes back and forth against the two walls and the current density can be used to identify and characterize the velocity of the generated wave fronts. For such a purpose, the rich oscillatory features of the current that reside in the shock front can be ignored. Here, however, we examine those features explicitly and focus on the insights they provide into our understanding of the fluid dynamics in dispersive shock scenarios. Moreover, we highlight the fact that whilst this discussion takes place in the context of ultra-cold Bose gases, a broad range of physical systems and dynamical situations are known to produce these types of shock waves \cite{Zakharov1968,Taylor1970,Tran1977,Mo2013,Rolley2007,Dominici2015,Rothenberg1989,Wan2007,Davis2009,Crosta2012,El2007,Carusotto2006,Kamchatnov2006,Kamchatnov2007,Hoefer2016,Hoefer2017,Hoefer2008,Engels2009,El2009,Engels2008}
(which have been recognized for their fundamental and ubiquitous nature \cite{Hoefer2016}), and any system governed by the nonlinear Schr\"odinger equation or GPE is likely to benefit from a number of the insights we obtain here as well.

\section{Model}
We start by recalling the Lieb-Liniger Hamiltonian for a uniform 1D Bose gas, in the second quantized form, describing a system of $N$ particles free to move along a 1D ring (i.e., with periodic boundary conditions) of length $L$ and  interacting via repulsive contact interactions \cite{Lieb-Liniger1963}:
\begin{align}
	\hat{H}_{\text{1D}}	=&-\frac{\hbar^{2}}{2m}  \int dx\, \hat{\Psi}^{\dagger}(x) \frac{\partial^{2}}{\partial x^{2}} \hat{\Psi}(x)\nonumber\\
	&+ \frac{g_{\mathrm{1D}}}{2} \int dx\, \hat{\Psi}^{\dagger}(x) \hat{\Psi}^{\dagger}(x) \hat{\Psi}(x) \hat{\Psi}(x). \label{eq:H s-wave 1D_CurrentCHAP}
\end{align}
Here, $\hat{\Psi}(x)$ is the bosonic field operator, $m$ is the mass of the particles, and $g_{\mathrm{1D}}$ is the interaction strength, with $g_{\mathrm{1D}}>0$ for repulsive interactions. Away from confinement induced resonances, the interaction strength is given by $g_{\mathrm{1D}}\simeq2\hbar a\omega_{\perp}$ \cite{Olshanii1998}, where $a$ is the three-dimensional $s$-wave scattering length and $\omega_{\perp}$ is the frequency of the transverse trapping potential, which is assumed harmonic.

We study shock waves that are generated from an initial density bump or dip on top of a non-zero uniform background. During evolution, the initial density gradients steepen, and a shock front forms. We monitor the fate of this shock front at a range of interaction strengths and initial equilibrium temperatures. In addition to analyzing the evolution of the mean particle number density 
\begin{equation}
\rho(x,t)=\langle \Psi^{\dagger}(x,t)\Psi(x,t)\rangle,
\label{eq:mean_density}
\end{equation} 
as in our previous work Ref.~\cite{Simmons2020}, here we also present and discuss the results for the mean current density, 
\begin{align}
    j(x,t) &= \frac{\hbar}{2mi} \left<\hat{\Psi}^{\dagger}(x,t)\frac{\partial \hat{\Psi}(x,t)}{\partial x} - \hat{\Psi}(x,t)\frac{\partial \hat{\Psi}^{\dagger}(x,t)}{\partial x}\right>.\label{eq:j_general}
\end{align}
The current density provides a useful probe in shock wave dynamics, since it offers a measure of the flow of particles and allows one to discern the motion of particles inside the shock region. The local sign of the current determines the direction of density flow in that region, where a positive (negative) sign indicates that particles are moving to the right (left) and zero current indicates that there is no net flow at that position in the fluid.

In all shock wave scenarios simulated below, the dynamics are initiated from an initial ($t=0$) density profile of the form
\begin{align}
    \rho(x,0)=N_{\mathrm{bg}}(1+\beta e^{-x^2/2\sigma^2})^{2}/L,
    \label{eq:rho_initial}
\end{align}
Such a density profile can be prepared prior to time $t=0$, either exactly or approximately \cite{initial_V}, as the ground state (at zero temperature $T=0$) or a thermal equilibrium state (at nonzero temperature $T\neq 0$) of a suitably chosen trapping potential. The trapping potential is then suddenly removed at time $t=0$, and the system with the above initial density profile is evolved under the Lieb-Liniger Hamiltonian, i.e., in a uniform potential of length $L$ with periodic boundary conditions. The above initial condition in the superfluid hydrodynamic description (see below) also assumes a zero initial current density $j_0(x,0)=0$, or equivalently a zero initial velocity field $v_0(x,0)$. In the mean-field and ideal Bose gas descriptions, on the other hand, we assume that the initial wave functions are given by $\Psi_0(x,0)=\Psi(x,0)=\sqrt{\rho(x,0)}$ (i.e., we assume that they are real-valued, without loss of generality). In all cases, $\sigma$ in Eq.~\eqref{eq:rho_initial}
 controls the width of the density profile and $N_{\mathrm{bg}}\!=\!|\psi_{\mathrm{bg}}|^{2}L\!=\!\rho_{\mathrm{bg}}L$ denotes the number of particles in the homogeneous background, related to the total number of particles via $N\!=\!N_{\mathrm{bg}}\big( 1+ \frac{\sqrt{\pi} \beta \sigma}{L} [ \beta\, \mathrm{erf}(\frac{L}{2\sigma}) +2 \sqrt{2}  \,\mathrm{erf} (\frac{L}{2\sqrt{2}\sigma}   )  ] \big)$. Furthermore, $\beta$ provides the height or depth  of the initial density perturbation, and we place no restriction on the sign of $\beta$ so as to explore both density bump situations where $\beta>0$, and additionally density dip situations where $-1<\beta<0$.

Within the Lieb-liniger model, a uniform system can be characterized by the dimensionless interaction parameter $\gamma\!=\!mg_{\mathrm{1D}}/\hbar^{2}\rho$, which provides the ratio of interaction energy to kinetic energy within the system. In the non-uniform density bump and dip scenarios under consideration, we choose to characterize the initial state of the gas using the value 
\begin{equation}
\gamma_{\mathrm{bg}}=\frac{mg_{\mathrm{1D}}}{\hbar^{2}\rho_{\mathrm{bg}}},
\end{equation} 
at the background density $\rho_{\mathrm{bg}}$.

The weakly interacting regime of the 1D Bose gas is realized when $\gamma_{\mathrm{bg}}\ll1$ \cite{Lieb-Liniger1963}, and at zero temperature ($T=0$) the dynamics of this system can be well approximated by the Gross-Pitaevskii equation (GPE) \cite{pitaevskii2016book}, 
\begin{equation}
    i\hbar\frac{\partial}{\partial t}\Psi_{0} = \left( -\frac{\hbar^2}{2m}\frac{\partial^{2}}{\partial x^{2}} + g_{\text{1D}}|\Psi_{0}|^2\right)\Psi_{0}, \label{eq:td-GPE}
\end{equation}
for the complex mean-field amplitude $\Psi_{0}(x,t)$, given by $\Psi_{0}(x,t)=\langle \hat{\Psi}(x,t)\rangle$ in the spontaneously broken symmetry approach.

Performing a Madelung transformation on the GPE to a set of real valued density and phase variables, $\Psi_{0}(x,t)\!=\!\sqrt{\rho_{0}(x,t)}e^{iS_{0}(x,t)}$, and then defining the fluid velocity via $v_{0}(x,t)\!=\!\frac{\hbar}{m}\frac{\partial}{\partial x} S_{0}(x,t)$, leads to the superfluid hydrodynamic equations
\begin{align}
    \frac{\partial \rho_{0}}{\partial t} &= -\frac{\partial}{\partial x}(\rho_{0} v_{0}), \label{eq:q-hydro-a}\\
    \frac{\partial v_{0}}{\partial t} &= -\frac{\partial}{\partial x}\left(\frac{1}{2}v_{0}^2+\frac{g_{\mathrm{1D}}\rho_{0}}{m}-\frac{\hbar^{2}}{2m^{2}}\frac{1}{\sqrt{\rho_{0}}}\frac{\partial^{2}}{\partial x^{2}} \sqrt{\rho_{0}} \right). \label{eq:q-hydro-b}
\end{align}
While these equations are written in terms of the fluid density and velocity, it is much more natural from a hydrodynamic perspective to consider equations for the conserved `charges' of the system. For a Bose gas at zero temperature one would consider these to be the fluid density $\rho_{0}(x,t)$ and the fluid `momentum' or current density $j_{0}(x,t)$. Hence, the current density arises naturally in the \textit{conservative} form of the superfluid hydrodynamic equations which are given by
\begin{align}
	\frac{\partial\rho_{0}}{\partial t} & =-\frac{\partial}{\partial x} j_{0},\label{eq:QHD-rho_j}\\
	\frac{\partial j_{0}}{\partial t} & =-\frac{\partial}{\partial x}\left[\frac{1}{\rho_{0}} j_{0}^{2} +\frac{P}{m}\right] + \rho_{0} \frac{\partial}{\partial x} \left[\frac{\hbar^{2}}{2m^{2}\sqrt{\rho_{0}}}\laplacian\sqrt{\rho_{0}}\right],\label{eq:QHD-j}
\end{align}
where the current density $j_{0}(x,t)$ \cite{current_density_vs_probability_current} and the pressure of the gas $P(x,t)$ are given by
\begin{align}
    j_{0} = \rho_{0}v_{0} = \frac{\hbar}{2mi} \left(\Psi_{0}^{*} \frac{\partial \Psi_{0}}{\partial x} - \Psi_{0} \frac{\partial \Psi_{0}^{*}}{\partial x}\right),
    \label{eq:j_0}
\end{align}
and
\begin{align}
    P(x,t) = \frac{g_{\mathrm{1D}}\rho_{0}(x,t)^{2}}{2},
\end{align}
respectively.

The last term in Eq. \eqref{eq:QHD-j} represents the so-called quantum pressure, which is responsible for the formation of the interference patterns that are produced in quantum shock waves \cite{Simmons2020}. This term does not behave like the usual pressure $P(x,t)$ but rather it acts like a source term or effective Bohm potential \cite{Bohm1,*Bohm2} in the dynamics of the Bose gas, generating forces based on spacial variations in the density.

We note that Eq. \eqref{eq:j_general} simplifies to the mean-field expression $j_0=\rho_0v_0$ of Eq. \eqref{eq:j_0} under the spontaneous symmetry breaking assumption $\langle\hat{\Psi}\rangle=\Psi_{0}$. In the ideal Bose gas regime (see below) one uses the same expression of Eq. \eqref{eq:j_0} but with the mean-field amplitude $\Psi_{0}(x,t)$ replaced by the actual Schr\"{o}dinger wave function $\Psi(x,t)$. Expressions to compute the current density within the other approaches we employ in this work can be found in their respective appendices. Hereafter we denote the current density and particle number density within any approach simply as $j$ and $\rho$ respectively, with the understanding that different expressions apply depending on the interaction regime and approach used to calculate these quantities.

\section{Density bump scenarios}
We begin our analysis by examining situations where $\beta>0$, i.e., for scenarios with an initial density bump that then expands into the non-zero background during evolution.

\subsection{Ideal Bose gas and weakly interacting regimes}

In the ideal (noninteracting) Bose gas regime at $T=0$, the dynamics are governed by the standard time-dependent Schr\"{o}dinger equation for the wave function evolving from the ground-state $\Psi(x,0)=\sqrt{\rho(x,0)}=\psi_{\mathrm{bg}}(1+\beta e^{-x^2/2\sigma^2})$. In the single-particle case, the wave function is normalized to one, whereas in the many-particle case, considered here, it is normalized to $N$, and therefore $|\psi_{\mathrm{bg}}|^2=N_{\mathrm{bg}}$ in order to give the initial density profile of Eq.~\eqref{eq:rho_initial}. The time-dependent solution to the Schr\"{o}dinger equation for this problem can be derived analytically and is given by \cite{Simmons2020}
 \begin{equation}
    \Psi(x,t)=\psi_{\mathrm{bg}}\left(1+\frac{\beta\sigma}{\sqrt{\sigma^2+i\hbar t/m}} e^{-x^2/2(\sigma^2+i\hbar t/m)}\right),
\label{eq:initial_wf}
\end{equation}
or alternatively $\Psi(x,t)\!=\!\psi_{\mathrm{bg}}[1\!+\!B(x,t)e^{i\varphi(x,t)}]$ with amplitude
\begin{align}
    B(x,t) \equiv \frac{\beta \sigma} {[\sigma^4+\hbar^2 t^2/m^2]^{1/4}}e^{-x^2\sigma^2/2[\sigma^4+\hbar^2 t^2/m^2]} \label{eq:beta_currents}
\end{align}
and phase
\begin{align}
    \varphi(x,t) \equiv \frac{\hbar t x^2}{2m[\sigma^4+\hbar^2 t^2/m^2]} - \frac{1}{2}\arctan\left(\frac{ \hbar^2 t^2}{m^2\sigma^4}\right). \label{eq:varphi_currents}
\end{align}
This allows for the density of the gas to be written as
\begin{align}
    \rho(x,t) &= |\Psi(x,t)|^2 \nonumber\\
    &= N_{\mathrm{bg}}[1 + B(x,t)^2 + 2B(x,t)\cos\varphi(x,t)]/L, 
    \label{eq:ideal_rho_interference}
\end{align}
where it becomes clear from the form of Eq. \eqref{eq:ideal_rho_interference} that the oscillations which arise dynamically are the result of quantum mechanical interference.

Furthermore, one can determine that the current density in the ideal gas 
regime, $j(x,t)=\rho(x,t)v(x,t)$, is given by (see Appendix \ref{appendix:Probability current for an ideal Bose gas})
\begin{align}
	j(x,t) = &\frac{\hbar N_{\mathrm{bg}}}{m L} \left\{B'(x,t)\sin\varphi(x,t) \right.\nonumber\\
	+& \left.\varphi'(x,t)\left[ B(x,t)\cos\varphi(x,t) + B^{2}(x,t)\right]\right\}, \label{eq:ideal_current}
\end{align}
with
\begin{align}
	B'(x,t) &= \frac{\partial  B(x,t)}{\partial x}
	= -\frac{\beta \sigma^{3} x}{\left(\sigma^{4} + \hbar^{2} t^{2}/m^{2}\right)^{5/4}} \nonumber\\
	&\qquad\qquad\qquad\qquad\times e^{-x^{2} \sigma^{2}/2\left(\sigma^{4} + \hbar^{2} t^{2}/m^{2}\right)},\label{eq:B'}\\
	\varphi'(x,t) &= \frac{\partial \varphi(x,t)}{\partial x} 
	=  \frac{\hbar t x}{m \left(\sigma^{4} + \hbar^{2} t^{2}/m^{2}\right)}.\label{eq:varphi'}
\end{align}

\begin{figure}[tbp]
    \centering
    \includegraphics[width=8.5cm]{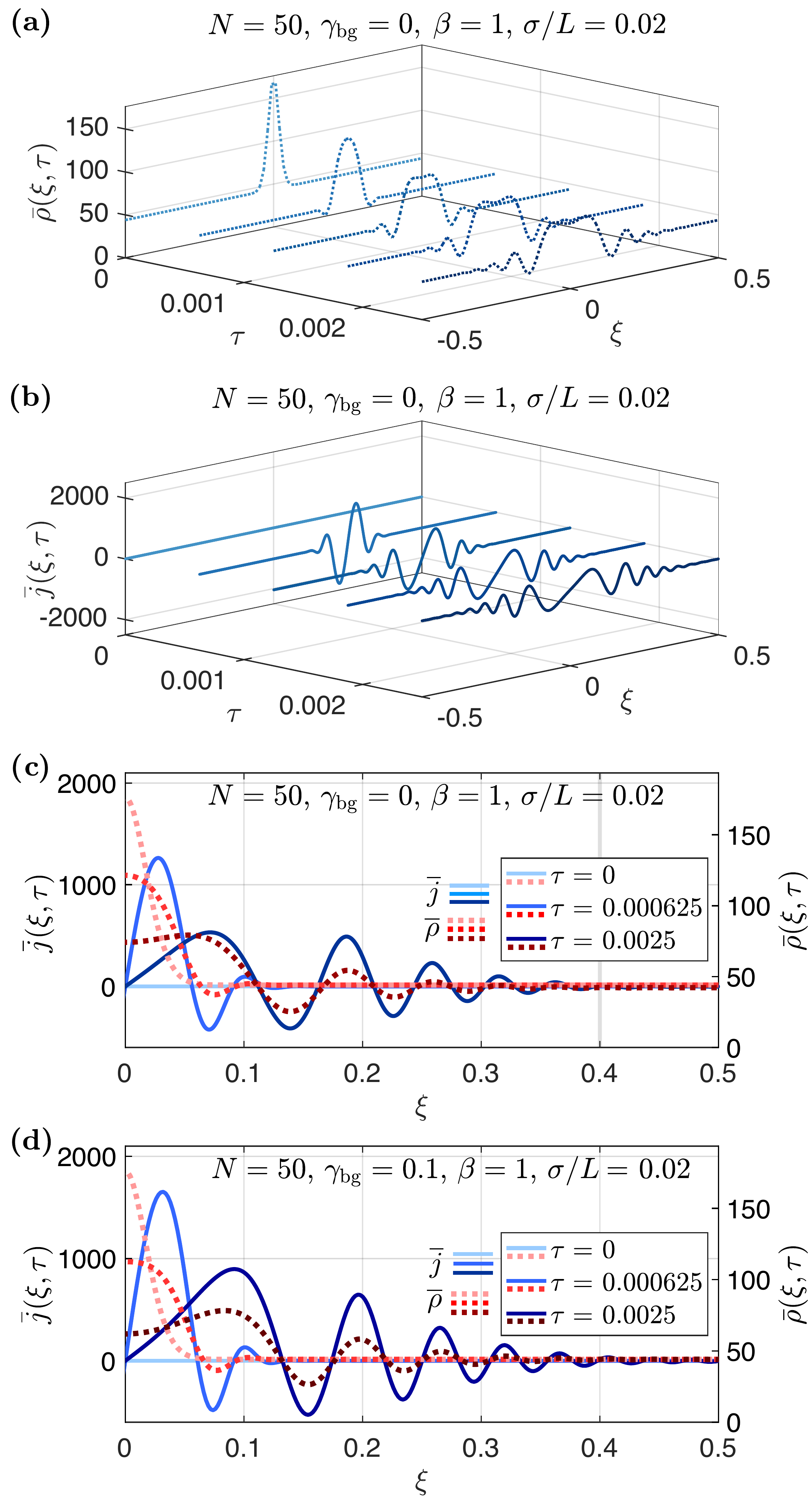}    
    \caption{Shock waves generated (from an initial density bump) in an ideal and weakly-interacting 1D Bose gas at zero-temperature ($T=0$). The gas begins with density profile $\rho=N_{\mathrm{bg}}(1+\beta e^{-x^{2}/2\sigma^{2}})^{2}/L$ and evolves in a box of length $L$ with periodic boundary conditions. (a) The dimensionless particle number density $\bar{\rho}=\rho L$ for an ideal gas ($\gamma_{\mathrm{bg}}=0$) of $N=50$ particles, with $\beta=1$ and $\sigma/L=0.02$ (leading to $N_{\mathrm{bg}}\simeq44.03$), at dimensionless times $\tau=t\hbar/mL^{2}$. The dimensionless position $\xi$ is scaled by the box length $L$ so that $\xi=x/L$. (b) The dimensionless current density $\bar{j}=jmL^{2}/\hbar$ corresponding to the same situation and times as in (a). (c) Both the dimensionless number density (dotted lines) and current density (solid lines) for select times of (a) and (b). While (a) and (b) provide a clear visualization of the shock wave scenario, hereafter we display results as in (c) to reduce space and provide clarity of detail with the recognition that the number density is an even function and current density an odd function. (d) The same as in (c) but now for a weakly interacting Bose gas with $\gamma_{\mathrm{bg}}=0.1$.}
    \label{fig:Currents_N50Bump_Ideal_Weakly}
\end{figure}

In Fig. \ref{fig:Currents_N50Bump_Ideal_Weakly}\,(a) we show the dynamics of the dimensionless density $\bar{\rho}(\xi,\tau)=\rho(x,t)L$ (where the dimensionless position and time are defined, respectively, according to $\xi=x/L$ and $\tau=t\hbar/mL^2$) and in (b) the dimensionless current $\bar{j}(\xi,\tau)=j(x,t)\,mL^{2}/\hbar$ of a shock wave generated in an ideal Bose gas, with $\gamma_{\mathrm{bg}}=0$ and a total of $N=50$ particles (where $N=50$ is chosen for comparison with other interaction regimes; see below). For clarity of detail and ease of interpretation we hereafter plot only the region $\xi=x/L\geq0$ at select dimensionless times $\tau$, with the recognition that the particle number density $\rho(x,t)$ is an even function $\rho(x,t)\!=\!\rho(-x,t)$, whereas the current density $j(x,t)$ is an odd function $j(x,t)\!=\!-j(-x,t)$ of $x$. Accordingly, in Fig. \ref{fig:Currents_N50Bump_Ideal_Weakly}\,(c) we re-plot some of the data in (a) and (b), where it becomes more clear that the oscillations, which arise in the shock wave train, have a period that is on the order of $\sim\!2\sigma$ \cite{Simmons2020}. We also observe a tell-tail sign that reinforces our interpretation from Ref.~\cite{Simmons2020} of the shock wave oscillations as a self-interference pattern: adjacent oscillations in the current density are commensurate with those in the particle number density and the oscillations alternate in sign. This indicates there are local counter-flows throughout the fluid, where for one interference fringe (any of the local density peaks) there is particle flow to the right, and for the next (any of the local density troughs) there is particle flow to the left, and so on. This implies there are counter-propagating fluid flows, which is similar to the mechanism of matter-wave interference occurring in counter-propagating BECs set by Bragg pulses (see, e.g., \cite{band2008interference}).
Overall, however, there is a net positive flow of particles to the right for $\xi\geq0$, indicative of the expansion of the initial bump into the background.

\begin{figure}[tbp]
    \centering
     \includegraphics[width=8.4cm]{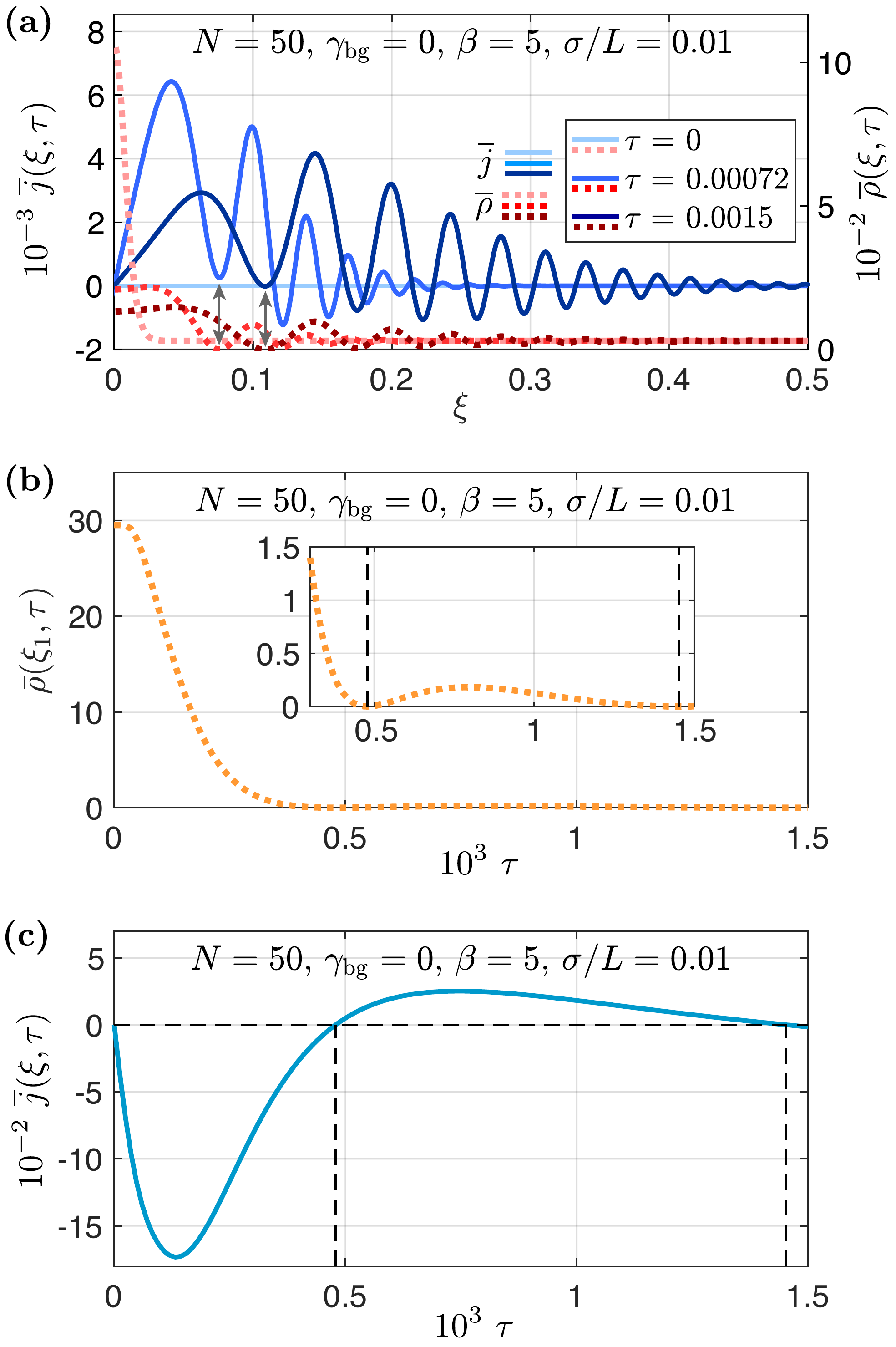}
    \caption{A vacuum point scenario in an ideal 1D Bose gas of $N=50$ particles at $T=0$, for $\beta=5$ and $\sigma/L=0.01$. Here, one of the number density fringes touches zero density during the evolution, due to a sufficiently large initial bump height $\beta$. (a) The dimensionless number density (dotted lines) and current density (solid lines) at different times $\tau$. The appearance of a vacuum point in the number density is marked by the disappearance (or reappearance) of sign-alternation in the current density at that location. We refer to this location as $\xi_{1}$ and it is denoted by the arrowheads in (a). 
    (b) Magnitude of the minimum in the trailing number density fringe over the course of the dynamics. This fringe touches zero twice during the dynamics and the times at which this occurs can be determined from Eq. \eqref{eq:beta_cr}. (c) Magnitude of the minimum in the trailing current density fringe over the course of the dynamics. The vertical dashed lines in (b) and (c) show the times at which a vacuum point occurs at the trailing interference fringe. The horizontal dashed line in (c) helps guide the eye to highlight where $\bar{j}(\xi_{1},\tau)$ changes sign, i.e., at the same time as $\bar{\rho}(\xi_{1},\tau)=0$. }    \label{fig:Currents_N50_VacuumPoint2}
\end{figure}

In Fig. \ref{fig:Currents_N50Bump_Ideal_Weakly}\,(d) we move into the weakly-interacting regime (with $\gamma_{\mathrm{bg}}=0.1$ and the same $N=50$ as before), where the solution is given by the numerical integration of the mean-field GPE or superfluid hydrodynamic equations. The dynamics of the dissolution of the initial density bump here is similar to the ideal Bose gas case; however, we point out that the interactions cause an overall increase in the strength of the currents produced and that the initial bump begins to separate into two distinct traveling wave packets in this case (one to the left, not shown, and one to the right, shown), indicated by the decrease in the density at $x=0$ closer to the background level. This is consistent with the expectation that the now non-zero pressure $P=g_{\mathrm{1D}}\rho_{0}^{2}/2$ is helping to push particles out of any local density bumps. In addition, we note that the trailing edge of the shock wave now travels at the speed of sound in the background $c_{\mathrm{bg}}=\sqrt{g_{\mathrm{1D}}\rho_{\mathrm{bg}}/m}$ (whereas in the $T=0$ ideal Bose gas, the speed of sound was zero).

We next investigate situations where the initial bump height $\beta$ is comparatively large ($\beta\gg 1$), so that the initial peak density $\rho(0,0)$ is much larger than the background density $\rho_{\mathrm{bg}}$. For the sake of analytical insight we present results again in the ideal Bose gas regime and note that the situation remains very similar for weak interactions. In the case of large $\beta$, we investigate a so-called \textit{vacuum point} scenario \cite{El1995} where the density of an interference fringe touches zero during evolution and the fluid velocity becomes undefined at that location. Despite the singularity that occurs at the vacuum point in the fluid velocity, the current density (or fluid momentum) at this location remains finite and is well defined.

The vacuum point  scenario is well known to the hydrodynamic community in the weakly-dispersive Thomas-Fermi regime of a weakly-interacting 1D Bose gas \cite{El1995,Cornell2006,El2007,Engels2008,Hoefer2016,Hoefer2017,Kamchatnov2021}, yet we are unaware of its identification in the non-interacting, strongly-dispersive ideal Bose gas regime. Due to the analytics we have readily available here, it is possible to determine the minimum or critical height $\beta_{\mathrm{cr}}$ required for the vacuum point to occur at time $t_{\mathrm{vac}}$ during evolution, and this depends on the initial bump width $\sigma$. By considering the lower envelope of the density \eqref{eq:ideal_rho_interference} one finds that the critical height can be determined using (see Appendix \ref{appendix:Critical height for a vacuum point to occur}),
\begin{align}
	\beta_{\mathrm{cr}}^{(n)} &= \frac{1}{\sigma} \left(\sigma^{4} + \hbar^{2} t_{\mathrm{vac}}^{2}/m^{2}\right)^{1/4} \nonumber\\
	&\qquad\qquad\times e^{\frac{m \sigma^{2}}{\hbar t_{\mathrm{vac}}} \left[\pi\left(2n-1\right) + \frac{1}{2}\arctan(\frac{\hbar^{2} t_{\mathrm{vac}}^{2}}{m^{2} \sigma^{4}})\right]} \label{eq:beta_cr}
\end{align}
where $n=1,2,3,...$ denotes the fringe number (increasing from the trailing to leading edge) that will become a vacuum point, and $t_{\mathrm{vac}}$ the time(s) that this will occur.

As opposed to the case for a step-like initial density profile, which is most often considered in the literature \cite{El1995,Cornell2006,Hoefer2016}, for the initial (raised) Gaussian squared profile that we consider here it is possible for the vacuum point to occur more than once at a given interference fringe during the system's evolution. The times at which the vacuum point occurs are marked by the disappearance (or reappearance) of sign-alternation in the corresponding fringe of the current density $j(x,t)$. Figure~\ref{fig:Currents_N50_VacuumPoint2}\,(a) provides snapshots of the density and current for an example where a vacuum point occurs twice at the trailing ($n=1$) interference fringe. Figs. \ref{fig:Currents_N50_VacuumPoint2}\,(b) and (c) trace the time evolution of the corresponding heights at the minimum of the trailing interference fringe, whose spatial location we label with $\xi_{1}(\tau)$. For the parameters used here, $\beta=5$ and $\sigma/L=0.01$, Eq. \eqref{eq:beta_cr} predicts that the vacuum point will occur at the trailing fringe at dimensionless times $\tau_{\mathrm{vac}}=t_{\mathrm{vac}}\hbar/mL^{2}\simeq0.479\times10^{-3}$ and $1.45\times10^{-3}$, which are marked in Figs \ref{fig:Currents_N50_VacuumPoint2}\,(b) and (c) with vertical dashed lines. At these times, the current at $\xi_{1}(\tau)$ changes sign, providing a transient regime between the occurrences of the vacuum points where the current is no longer sign-alternating for that interference fringe.

\subsection{Weakly interacting Bose gas in the Thomas-Fermi limit}

Considering now a much larger number of particles in the system ($N=2000$) at $\gamma_{\mathrm{bg}}=0.01$ brings us into the Thomas-Fermi regime of the weakly interacting 1D Bose gas (i.e., where the mean-field interaction energy of the particles dominates their kinetic energy). In this regime, we start by exploring the shock wave densities and currents that arise in the previously examined situation of Fig. 1\,(c) from Ref.~\cite{Simmons2020} with $\beta=1$, and then investigate those that emerge in a weakly-dispersive vacuum point scenario for $\beta=5$. In both cases we complement the mean-field solution with results from the truncated Wigner approach (which we initialize as the Bogoliubov ground-state of a suitably chosen trapping potential, see Appendix \ref{appendix:The truncated Wigner approach_currents}) and the classical-field stochastic projected Gross-Pitaevskii equation (SPGPE, see Appendix \ref{appendix:SPGPE}). These alternative approaches go beyond the mean-field GPE description and allow us to discuss, respectively, the effect of quantum fluctuations and finite temperature in these scenarios.

\begin{widetext}
\begin{figure*}[tbp]
    \centering
    \includegraphics[width=0.97\linewidth]{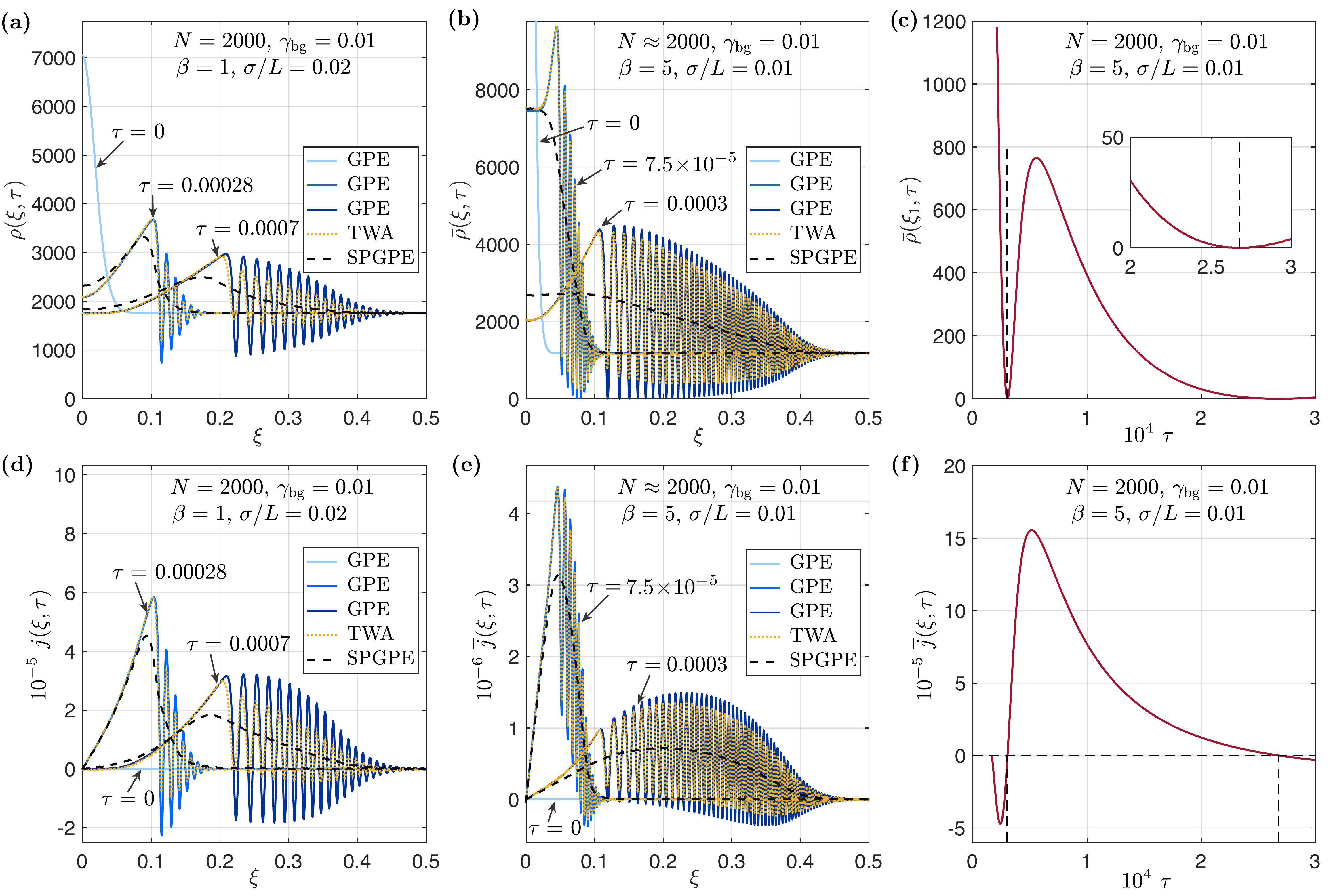}
    \caption{Shock waves generated (from an initial density bump) in the Thomas-Fermi regime of a 1D Bose gas. Blue  (shades of gray, at different times) lines indicate the mean-field solution, yellow (light gray) dotted lines indicate the results of the truncated Wigner approach (TWA, see Appendix \ref{appendix:The truncated Wigner approach_currents} for details) and black dashed lines provide the prediction of the SPGPE for which the initial state is characterized by a dimensionless temperature of $\overline{\mathcal{T}}=T/T_{d}$, where $T_{d}=\hbar^{2}\rho_{\mathrm{bg}}^{2}/2mk_{B}$ is the temperature of quantum degeneracy at the background density (see Appendix \ref{appendix:SPGPE} for further details). Panels (a) and (d) show the particle number density and current density, respectively, for the situation with $\beta=1$ and $\sigma/L=0.02$ explored previously in Ref. \cite{Simmons2020}; the SPGPE results shown in these panels are for $\overline{\mathcal{T}}=0.05$. Panels (b) and (e) show the same, but for a situation with $\beta=5$ and $\sigma/L=0.01$ which causes a vacuum point to occur in the mean-field solution; the SPGPE results in this case are for $\overline{\mathcal{T}}=0.5$. In (b) the height of the initial profile is cropped so as to display more detail in the density at subsequent times. Panels (c) and (f) trace the respective height of the trailing interference fringe over the course of the mean-field GPE dynamics, which remains initially at the background value until the wave breaks and a dispersive shock occurs. The vertical dashed lines denote the times $\tau_{\mathrm{vac}}\approx0.30\times10^{-4}$ and $2.67\times10^{-4}$ that this fringe becomes a vacuum point (determined numerically), with the horizontal dashed line in (f) used to guide the eye. All TWA and SPGPE results presented in this work are ensemble averages over 100,000 stochastic trajectories.
    }    \label{fig:Currents_N2000_VacuumPoint2}
\end{figure*}
\end{widetext}

In the Thomas-Fermi regime, the pressure $P=g_{\mathrm{1D}}\rho^{2}/2$ of the gas is greater than in the nearly-ideal situation of \ref{fig:Currents_N50Bump_Ideal_Weakly}\,(d) and this causes the left (not shown) and right moving wave trains to separate from each other quickly, see Figs. \ref{fig:Currents_N2000_VacuumPoint2}\,(a) and (d) which are for a bump height of $\beta=1$. Moreover, the oscillations in this case have a period on the order of the healing length in the background $l_{\mathrm{h}}=\hbar/\sqrt{mg\rho_{\mathrm{bg}}}=L/N_{\mathrm{bg}}\sqrt{\gamma_{\mathrm{bg}}}$ \cite{Simmons2020}, which is now smaller than $\sigma$ and is the shortest relevant length-scale in the system. Here, the current affords a similar interpretation as in the preceding section, and the oscillations are sign-alternating and therefore produce counter-flows between adjacent peaks and troughs. Again, such an insight gained from the current density provides a reinforcement of the conclusion that these oscillations can be interpreted as interference fringes across the phase coherent gas.

Furthermore, Figs. \ref{fig:Currents_N2000_VacuumPoint2}\,(b) and \ref{fig:Currents_N2000_VacuumPoint2}\,(e), which are for a bump height of $\beta=5$, provide similarly the density and current for a scenario during which a vacuum point occurs. In this situation, the location of the vacuum point drifts (from left to right) into the middle of the shock wave during evolution and multiple interference fringes touch zero density throughout the mean-field dynamics. Moreover, for the initial Gaussian squared  profile we consider, the vacuum point eventually drifts back out of the shock front (from right to left) at longer times, and by the end of the simulation shown here (at $\tau=0.0003$), each fringe minimum that at some point became a vacuum point has re-touched zero density again. This behavior is captured in Figs. \ref{fig:Currents_N2000_VacuumPoint2}\, (c) and (f) which trace the minimum height, according to the mean-field prediction, of the trailing fringe in the density and current respectively. We see that this fringe minimum quickly becomes a vacuum point and returns to zero density again late in the dynamics. The times at which this interference fringe becomes a vacuum point can be determined numerically and are approximately $\tau_{\mathrm{vac}}\approx0.30\times10^{-4}$ and $2.67\times10^{-4}$. To the best of our knowledge, this is the first time such behavior of the vacuum point has been reported.

When one includes the effect of quantum fluctuations using the truncated Wigner approach (see yellow lines in Fig. \ref{fig:Currents_N2000_VacuumPoint2}), the visibility of the interference pattern (i.e., the amplitude of the density oscillations), along with the strength of the currents produced, is diminished compared with the mean-field GPE result. This reduction can be credited to shot-to-shot fluctuations in the position of the oscillations around their mean \cite{Simmons2020}.
Moreover, in Figs. \ref{fig:Currents_N2000_VacuumPoint2}\,(b) and (e) we see that the quantum fluctuations prevent an actual vacuum point from occurring, i.e., the mean density never touches zero. Hence we conclude that the vacuum point observed in the GPE is a strictly mean-field effect, i.e., it is an artifact of the mean-field approximation and it disappears once the effects of quantum fluctuations are taken into account \cite{Wigner_experimental_runs}. This was conjectured in Ref. \cite{El1995}; ``in the vicinity of the vacuum point the NLS-approximation can fail and a more complex model should be evolved". We have thus confirmed this conjecture here through the simulation of both the truncated Wigner and SPGPE models.

The effect of thermal fluctuations, that we simulated for finite temperature systems through the SPGPE approach, is similar to that of quantum fluctuations; see black dashed lines in Fig. \ref{fig:Currents_N2000_VacuumPoint2}, which are predictions of the SPGPE for dimensionless temperatures of $\overline{\mathcal{T}}=T/T_{d}=0.05$ and $0.5$ (see figure caption for details). The thermal fluctuations cause such a reduction in the visibility of the interference contrast, or in the strength of the currents which are produced, that the oscillatory wave train disappears entirely. As such, thermal fluctuations also prevent the vacuum point from occurring in the mean density, see Fig. \ref{fig:Currents_N2000_VacuumPoint2}\,(b). We point out, however, that  the vacuum point may still occur in individual stochastic trajectories, which themselves can be thought of as representing the outcomes of individual experimental runs \cite{Blakie2008}.

\subsection{Strongly interacting and Tonks-Girardeau regimes}
\begin{figure}[tbp]
    \centering
    \includegraphics[width=8.3cm]{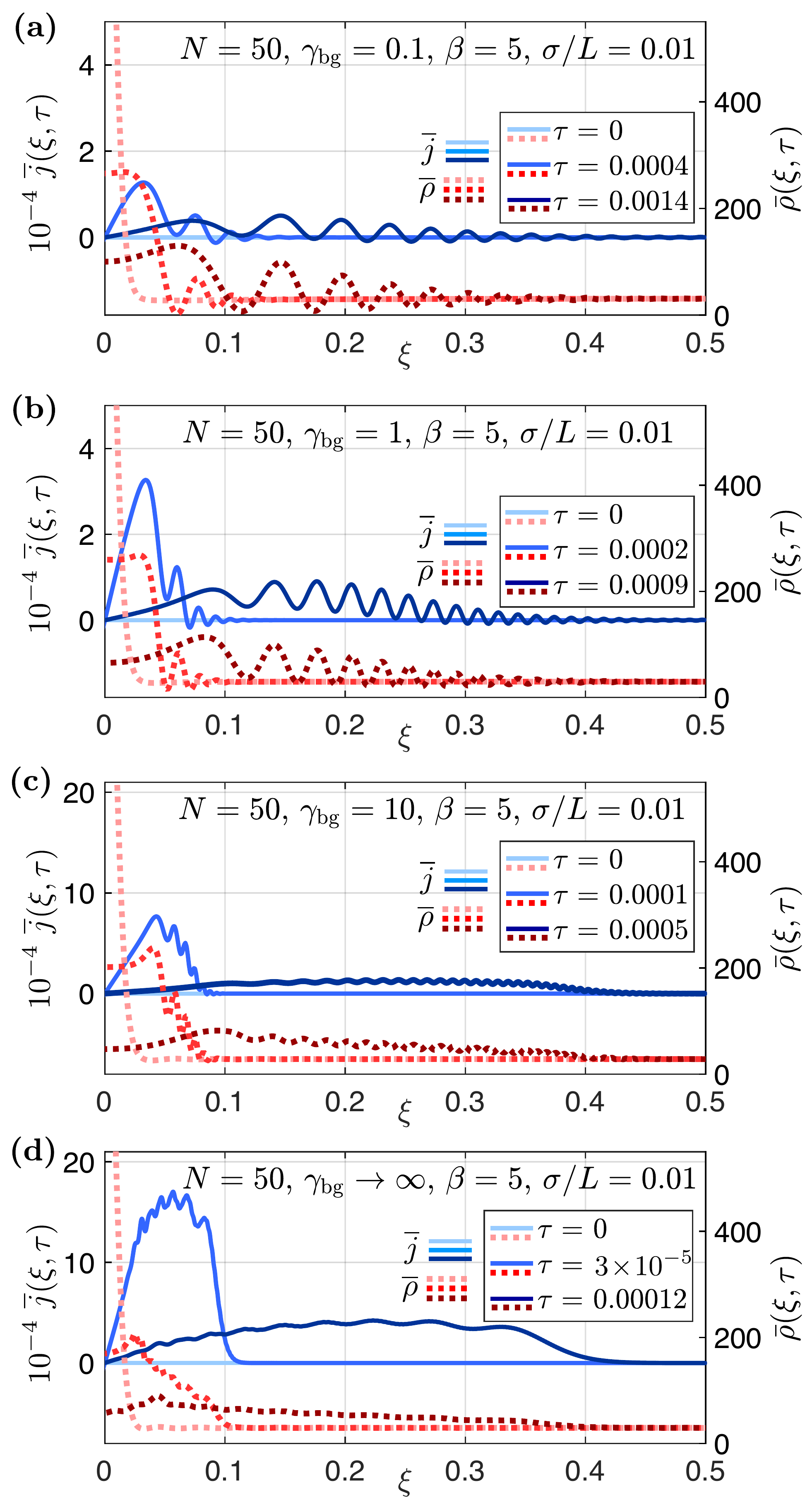}
    \caption{Shock waves generated (from an initial density bump) in a 1D Bose gas at weak ($\gamma_{\mathrm{bg}}=0.1$),  intermediate ($\gamma_{\mathrm{bg}}=1$), strong ($\gamma_{\mathrm{bg}}=10$), and infinitely strong ($\gamma_{\mathrm{bg}}\to \infty$) interactions. (a) through (d) show the particle number densities and currents for the same initial situation as in Fig. \ref{fig:Currents_N50_VacuumPoint2}\,(a) i.e. for $N=50$, $\beta=5$, and $\sigma/L=0.01$, except now for non-zero interaction strengths. The initial peak density in these plots is approximately $\overline{\rho}(0,\tau=0)\simeq 1060$, however, the vertical axis is cropped here to make the details more visible in the time-evolved densities at later times.
    }
    \label{fig:Currents_N50Bump_Interaction_Sweep}
\end{figure}

Here, we re-consider the scenario initiated from the same density profile as in Fig. \ref{fig:Currents_N50_VacuumPoint2}\,(a) for $N=50$ particles and $\beta=5$, but we now sweep the interaction strength into the weakly interacting $\gamma_{\mathrm{bg}}\ll 1$, intermediate $\gamma_{\mathrm{bg}}\sim1$, strongly $\gamma_{\mathrm{bg}}>1$, and Tonks-Girardeau (TG) $\gamma_{\mathrm{bg}}\rightarrow\infty$ regimes. 
The density and current in these situations are shown in Figs. \ref{fig:Currents_N50Bump_Interaction_Sweep}\,(a) through (d). Here, the solutions for $\gamma_{\mathrm{bg}}=0.1$, $\gamma_{\mathrm{bg}}=1$, and $\gamma_{\mathrm{bg}}=10$ are obtained using matrix product state methods (similarly to Ref. \cite{Simmons2020}, see Appendix VI of that work), whereas for $\gamma_{\mathrm{bg}}\rightarrow\infty$ we use Fermi-Bose mapping \cite{Girardeau1960,Yukalov2005} and exact diagonalization of the free fermions problem (see Appendix \ref{appendix:Exact diagonalisation in the TG limit}).

It has already been identified in Ref. \cite{Simmons2020}, and can be seen again here, that the interference contrast in the density disappears as the interaction strength is increased. This is due to a decrease in coherence length of the system, which reduces down to the size of the mean inter-particle separation $1/\rho_{\mathrm{bg}}$ in the TG limit. In this limit the initial state itself is not smooth but rather contains small-scale Friedel oscillations \cite{Friedel1958} whose period is on the order of $1/\rho_{\mathrm{bg}}$, and which are not visible initially on the scale plotted in Fig.~\ref{fig:Currents_N50Bump_Interaction_Sweep}\,(d). During evolution, the  oscillations that appear towards the front of the shock are simply deformations of the pre-existing Friedel oscillations, whereas the irregular peaks that develop dynamically toward the rear of the shock can be interpreted as Friedel-type oscillations between different branches of the Fermi momentum (in phase-space) \cite{Protopopov2013}. In either case, with the increase of interaction strength, these density modulations shift towards occurring predominantly \emph{above} the level of the background density rather than \emph{around} it, which is unlike the density oscillations in the dispersive shock-wave interference fringes. Through Fig. \ref{fig:Currents_N50Bump_Interaction_Sweep}\,(a) to (c), which are for $N=50$, we show that the crossover between interference and the lack thereof is easily identifiable by examining the oscillations in the current density, which lose their sign-alternating and large oscillatory nature at higher interaction strengths, indicating that there are no longer any counter-flows, 
or interference, in the fluid. All these features imply that any vacuum point that may have existed for weak or zero interaction under a certain combination of parameters would cease to exist as the interaction strength is increased.

\begin{figure}[tbp]
    \centering
    \includegraphics[width=8.3cm]{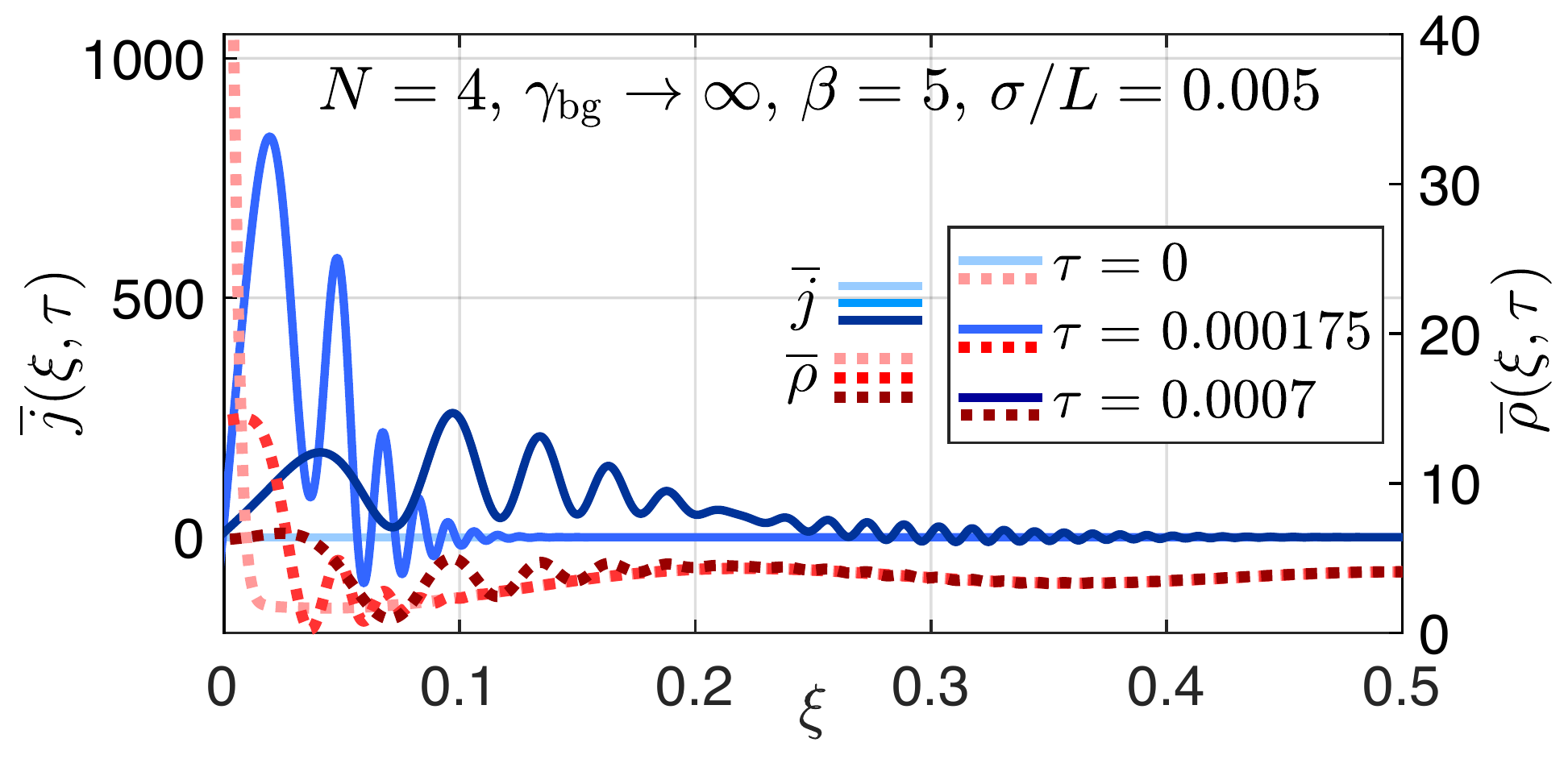}
    \caption{Shock waves generated (from an initial density bump) in a 1D Bose gas in the TG limit  ($\gamma_{\mathrm{bg}}\to \infty$), for $N=4$, $\beta=5$, and $\sigma/L=0.005$. In this scenario, the width of initial density bump $\sigma$ is much smaller than the period of pre-existing Friedel oscillations in the initial density profile. The initial peak density here is approximately $\overline{\rho}(0,\tau=0)\simeq 70$, however, the vertical axis is cropped to make the details more visible in the time-evolved density at later times. 
    }
    \label{fig:Currents_N4_Bump_TG}
\end{figure}

In the TG limit for $N=4$ (rather than $N=50$), when the initial bump is much narrower than the mean inter-particle separation ($\sigma\ll1/\rho_{\mathrm{bg}}$), see Fig. \ref{fig:Currents_N4_Bump_TG}, one again observes the emergence of interference fringes. They are, however, contained within a single Friedel oscillation period (which are visible in the initial state of this scenario) over which the gas is coherent. Indeed, the phase coherence length in the TG regime is on the order of the mean interparticle separation, $1/\rho_{\mathrm{bg}}$ \cite{Cazalilla2004}, which is also the characteristic period of Friedel oscillations \cite{Friedel1958}. As such, an initial bump with a width that is much smaller than $1/\rho_{\mathrm{bg}}$ expands into a locally phase coherent background---hence the interference fringes that we see in this regime. The current density in Fig. \ref{fig:Currents_N4_Bump_TG} at times $\tau=1.75\times 10^{-4}$ and $\tau=7\times10^{-4}$ shows clearly this containment of sign-alternating counter-flows to the order of the mean inter-particle separation. 
We thus conclude again that the interference fringes are phenomena produced when the local phase coherence length of the gas is much larger than the width of the initial bump.

\section{Density dip scenarios}
Shifting our focus to situations where $\beta<0$, the initial density profile now contains a dip, which the background density subsequently evolves to fill. Here we consider scenarios that mimic their density bump counterpart and choose whole total particle numbers $N$ which provide the closest value of $N_{\mathrm{bg}}$ to the respective density bump situation.

\subsection{From the ideal Bose gas to the strongly interacting regime}

\begin{figure}[tbp]
    \centering
    \includegraphics[width=8.3cm]{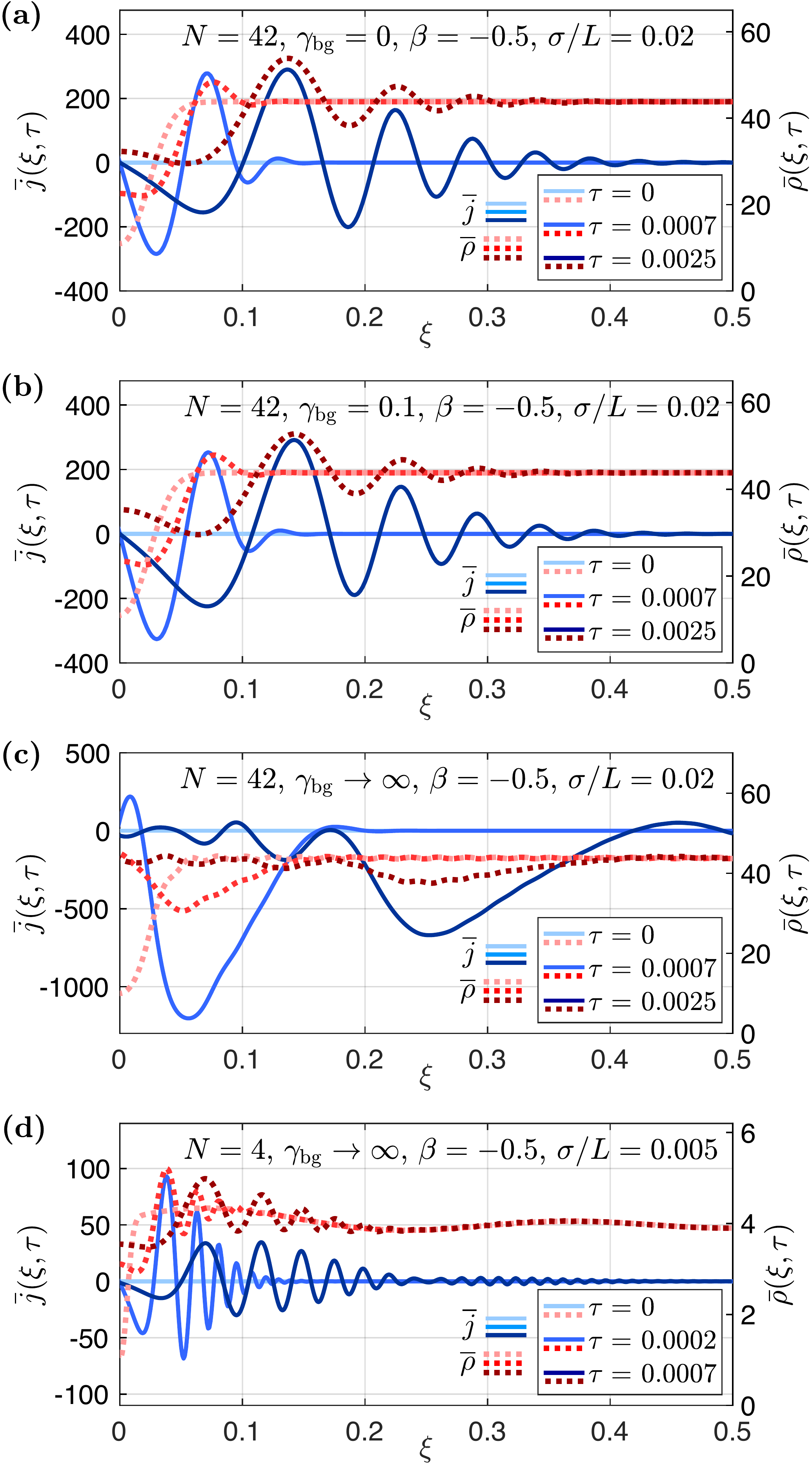}
    \caption{Shock waves generated (from an initial density dip) in a zero-temperature ($T=0$) 1D Bose gas for a range of interaction strengths. The particle number densities and current densities are shown for: (a) $N=42$, $\gamma_{\mathrm{bg}}=0$ [the density dip equivalent of Fig. \ref{fig:Currents_N50Bump_Ideal_Weakly}\,(c)]; (b) $N=42$, $\gamma_{\mathrm{bg}}=0.1$ [the density dip equivalent of Fig. \ref{fig:Currents_N50Bump_Ideal_Weakly}\,(d)]; (c) $N=42$, $\gamma_{\mathrm{bg}}\rightarrow\infty$ [the density dip equivalent of Fig. \ref{fig:Currents_N50Bump_Interaction_Sweep}\,(c)]; and (d) $N=4$, $\gamma_{\mathrm{bg}}\rightarrow\infty$ [the the density dip equivalent of Fig. \ref{fig:Currents_N50Bump_Interaction_Sweep}\,(d)].
    }
    \label{fig:Currents_N50Dip_Interaction_Sweep}
\end{figure}

We first explore the effect of inter-particle interactions for relatively small atom numbers ($N=42$) as we vary the interaction strength from the ideal Bose gas regime at $T=0$ through to the TG regime of infinitely strong interactions. The resulting snapshots of the density and current in these examples are shown  in Figs. \ref{fig:Currents_N50Dip_Interaction_Sweep}\, (a) through (c).

While density dip scenarios are usually well known for producing trains of gray solitons \cite{Dutton2001,Kamchatnov2002,El2006,Damski2006,Engels2009,Kamchatnov2021}, here in the ideal Bose gas regime (which does not support solitons) we see that this situation results purely in self-interference. Similar to the equivalent bump scenario of Fig. \ref{fig:Currents_N50Bump_Ideal_Weakly}\,(c), the current remains sign-alternating with counter-flows across the fluid, yet the sign of these oscillations themselves is flipped, indicating a change in direction of net particle flow which is now towards the center of the system such that the background is filling in the initial density dip.

In the regime of weak interactions the physical picture remains much the same as in the equivalent bump case Fig. \ref{fig:Currents_N50Bump_Ideal_Weakly}\,(d), where the oscillations in the density are still the result of quantum interference, and the initial dip, as it begins to fill, splits into two counter-propagating oscillatory waves due to a non-zero pressure $P=g_{\mathrm{1D}}\rho_{0}^{2}/2$.

As one reaches the infinitely-interacting TG limit, we observe small-scale Friedel oscillations in the initial state of this scenario \cite{Friedel1958} (see Fig. \ref{fig:Currents_N50Bump_Interaction_Sweep}\,(c)), but the large density ripples present for weak interactions disappear. In these density dip situations, the actual shock itself occurs on the inside (or left) edge of the propagating wave, where in this TG limit there emerges some intriguing large scale (wavelength) oscillations that span several Friedel oscillations, and whose corresponding oscillations in the current show weak sign-alternating counter-flows.

We note that for large particle numbers $N$ density ripples, that we refer to as Bettelheim-Glazman oscillations (see Appendix \ref{appendix:Bettelheim-Glazman oscillations in the TG limit} and \cite{Bettelheim2012,Glazman2019,Watson2022}), exist across the shock front; these larger wavelength oscillations span numerous Friedel oscillations and their frequency is chirped toward the center of the wave-packet. 

The oscillations which emerge here at small $N$, however, do not appear to posses this same style of chirp and it does not seem appropriate to identify them as Bettelheim-Glazman oscillations, which are known to only appear in the limit of large $N$ \cite{Bettelheim2012,Glazman2019}. In fact, their true origin is quite unclear; they emerge dynamically on the opposite side of the shock wave compared to the usual interference pattern generated in the weakly-interacting regime, and they posses a much more periodic nature compared with the usual density peaks that arise as a result of Friedel-type oscillations between different Fermi-momentum branches \cite{Protopopov2013}. Nevertheless, determining their true cause is beyond the scope of this paper and we leave the identification of their origin for future work.

For even smaller $N$, however, where the initial bump width is smaller than the mean inter-particle separation, as in Figs. \ref{fig:Currents_N50Dip_Interaction_Sweep}\,(d), we again observe interference fringes that are contained within single Friedel oscillations, where the sign of the alternating currents is similarly flipped compared with the equivalent bump case from Fig. \ref{fig:Currents_N50Bump_Interaction_Sweep}.

\subsection{Shedding of gray solitons in the weakly interacting Bose gas in the Thomas-Fermi limit}

Finally, we explore the fate of an initial density dip in the weakly-dispersive Thomas-Fermi regime, which is achieved for much larger values of $N$ than in previous subsection. It is in this limit that the generation of gray soliton trains is possible and we identify the role that the initial dip width plays in the production of these solitons. We first consider the mean-field GPE solution and the insights it has to offer, before discussing the effects of quantum and thermal fluctuations in these scenarios.

\begin{widetext}
\begin{figure*}[tbp]
    \centering
    \includegraphics[width=0.99\linewidth]{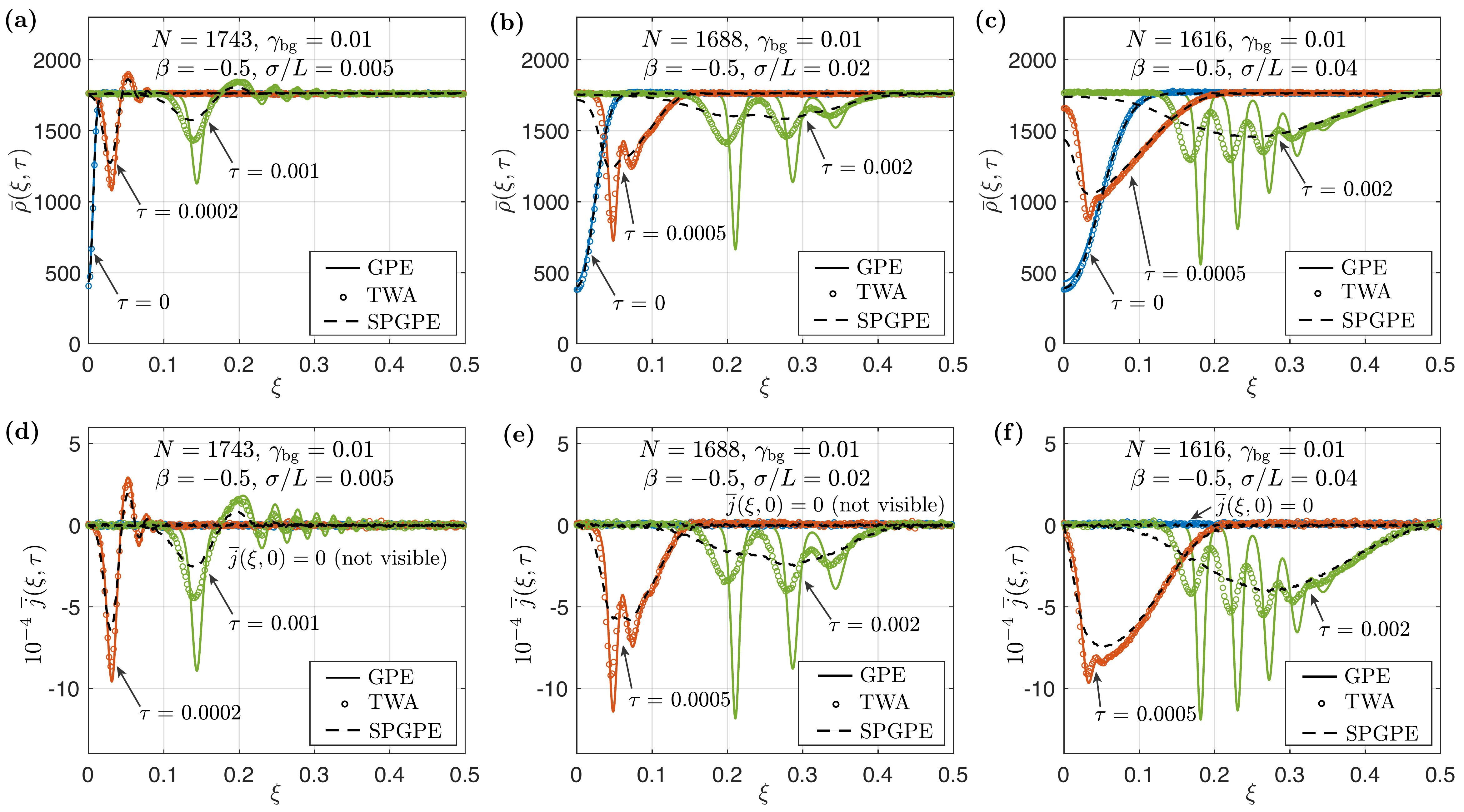}
    \caption{Particle number densities (top row) and currents (bottom row) that form from an initial density dip in a 1D Bose gas in the Thomas-Fermi regime. Solid lines represent the mean-field solution, open circles provide the result of the truncated Wigner approach (see Appendix \ref{appendix:The truncated Wigner approach_currents} for details) and black dashed lines indicate the prediction of the SPGPE at a dimensionless temperature of $\overline{\mathcal{T}}=T/T_{d}=0.01$ where $T_{d}=\hbar^{2}\rho_{\mathrm{bg}}^{2}/2mk_{B}$ is the temperature of quantum degeneracy at the background density (see Appendix \ref{appendix:SPGPE} for further details). The initial bump width increases from left to right where one generates more solitons for larger $\sigma$. In each case the total number of particles $N$ is chosen to be a whole number that gives the closest $N_{\mathrm{bg}}$ to that of Fig. \ref{fig:Currents_N2000_VacuumPoint2}\,(a); $N_{\mathrm{bg}}\simeq1761$ which leads to $l_{\mathrm{h}}/L\simeq0.0057$.}
    \label{fig:Currents_N2000_sigma_sweep}
\end{figure*}
\end{widetext}

For small widths, see Fig. \ref{fig:Currents_N2000_sigma_sweep}\,(a), where $\sigma/L=0.005$ is becoming comparable to the healing length $l_{\mathrm{h}}\simeq0.0057$, two gray solitons are shed from the initial density profile (one propagating in each direction) and a clearly identifiable interference chirp that precedes them in time is generated, which expands away from the solitons and reduces in visibility. These observations are reinforced by the current density in Fig. \ref{fig:Currents_N2000_sigma_sweep}\,(d) which shows sign-alternating currents in the interference region and a clear peak marking the location of the rightward moving soliton, whose propagation is supported due to particle flow towards the left.

As the dip width is increased and one goes deeper into the Thomas-Fermi regime, the interference pattern diminishes and the number of generated solitons grows. For a width $\sigma/L=0.02$, see Fig. \ref{fig:Currents_N2000_sigma_sweep}\,(b) [equivalent to the bump case of Fig. \ref{fig:Currents_N2000_VacuumPoint2}\,(a)], the initial dip sheds a total of six solitons with three traveling in each direction, and minimal interference is generated in front of the leading solitons, which can be confirmed from the current density in Fig. \ref{fig:Currents_N2000_sigma_sweep}\,(e). Due to their increasingly (to the right) shallower depths and therefore increasingly faster propagation velocities \cite{soliton_speed}, the solitons separate from each other as the system continues to evolve.

With twice larger width $\sigma/L=0.04$, see Fig. \ref{fig:Currents_N2000_sigma_sweep}\,(c), five or more solitons are shed on either side of the initial profile and interference fringes no longer appear in front of the soliton train. We note then that one can use the initial dip width $\sigma$ to precisely control and generate specific numbers of solitons on demand. On top of this, a sweep from interference fringes to controllable soliton generation can be achieved via any means that shift the system away from the strongly-dispersive ideal gas regime into the weakly-dispersive Thomas-Fermi limit, i.e., any change that decreases the healing length $l_{\mathrm{h}}=\hbar/\sqrt{mg_{\mathrm{1D}}\rho_{\mathrm{bg}}}=L/N_{\mathrm{bg}}\sqrt{\gamma_{\mathrm{bg}}}$ compared with $\sigma$. In particular, with a deeper initial profile (controlled by $\beta$) or increased $N_{\mathrm{bg}}$, or $\gamma_{\mathrm{bg}}$, it is possible to generate longer trains of numerous well separated solitons. These findings are consistent with results presented in Ref. \cite{Kamchatnov2021} that indicate the number of solitons generated from an initial density dip profile depends only on the initial distribution of the local speed of sound $c(x,0)=\sqrt{g_{\mathrm{1D}}\rho(x,0)/m}$. For clarity, in Appendix \ref{appendix:Identification of solitons in the mean-field GPE} we provide details around the identification of solitons in the mean-field GPE solution.

When quantum fluctuations are introduced (see open circles in each panel of Fig. \ref{fig:Currents_N2000_sigma_sweep}), they act to weaken or destroy the large-amplitude features present in the mean-field solution. For the interference present in Figs. \ref{fig:Currents_N2000_sigma_sweep}\,(a) and (d) the situation is the same as for the density bump; the interference contrast is reduced due to a reduction in the strength of sign-alternating counter-flows in the fluid, which itself is due to shot-to-shot fluctuations in the position of oscillations about their mean. This same shot-to-shot averaging has a somewhat more dramatic effect on the solitons---namely, on reducing the amplitude and broadening the soliton width---to the extent that, without the mean-field solution, it would be very difficult to identify that the density dips in the truncated Wigner result were remnants of solitons from single shots. Furthermore, the effect of thermal fluctuations on these features, for an initially finite-temperature system, is similar yet even more pronounced (see black dashed lines in each panel of Fig. \ref{fig:Currents_N2000_sigma_sweep}). So much so that for the situations of Fig. \ref{fig:Currents_N2000_sigma_sweep}\,(b) and (e), and (c) and (f), there remains little trace of any local density dip features, and the SPGPE result provides almost a smoothed average of the truncated Wigner dips or oscillations.

\section{Summary}

In summary, we have explored quantum shock wave scenarios in 1D Bose gases for a broad range of initial density perturbations and interaction strengths, including those far outside the commonly studied weakly-dispersive Thomas-Fermi regime. Through the use of the current density $j(x,t)$, in addition to particle number density $\rho(x,t)$, we have consolidated further the interpretation and understanding of quantum interference and soliton generation in these systems. Moreover, the ubiquitous nature of the shocks that we consider allows other fields of physics to obtain a useful perspective on the interesting and rich dynamics of these scenarios.

For Bose gases in 
initial density bump situations, the dispersive shock waves are the result of quantum interference and their visibility is reduced as the phase coherence length of the gas decreases, i.e., for increasing interaction strengths or the introduction of quantum or thermal fluctuations \cite{Simmons2020}. Here, we have shown that this reduced visibility is accompanied by reduced sign-alternating currents across the fluid. Furthermore, we have identified the possibility of vacuum points in the strongly-dispersive ideal Bose gas regime and shown that such vacuum points can occur more than once at a given interference fringe, where the times that they occur are marked by the disappearance or reappearance  of sign-alternating currents at that location. In the weakly-dispersive Thomas-Fermi limit of a weakly interacting 1D Bose gas we have shown that quantum and thermal fluctuations reduce the visibility of interference fringes and prevent an actual vacuum point from occurring in the mean density.

In density dip situations, the generation of both interference and mean-field gray solitons is possible in the weakly interacting regime where $\gamma_{\mathrm{bg}}<1$. We have shown that the number of solitons produced in the Thomas-Fermi limit can be effectively controlled through the width and depth of the initial density dip. We have also shown that quantum and thermal fluctuations each diminish the visibility of the aforementioned phenomena so that the remnant (in the average density) of the mean-field soliton core becomes broader and shallower.

The generation of soliton trains also disappears as the interaction strength of the gas is increased into the strongly interacting regime, where, for large particle numbers, we instead observe the emergence of Bettelheim-Glazman oscillations \cite{Bettelheim2012,Glazman2019,Watson2022} on top of a broad density depletion, similar to those predicted for a density bump scenario. We note that these oscillations have a different origin to Friedel oscillations, as they occur on different (larger) lengthscales. In the examples presented here (see Appendix \ref{appendix:Bettelheim-Glazman oscillations in the TG limit}), which were for $N>800$, we have been able to discriminate between the Friedel oscillations, which have a period of $1/N$, and the Bettelheim-Glazman density ripples, which have a larger oscillation period. This was not possible in the prior work of Ref.~\cite{Simmons2020}, which treated a much smaller number of particles ($N = 50$) in the Tonks-Girardeau regime. We also note that Bettelheim-Glazman oscillations are chirped towards the center of the overall density depletion (which propagates away from the origin). This is different from the interference oscillations that occur in the weakly-interacting regime, which are chirped towards the front of the shock wave envelope. Overall, the mechanism behind the formation of Bettelheim-Glazman oscillations is yet to be understood, and we expect that such oscillations should be present also at very large but finite interaction strengths, i.e., away from the strictly TG limit of infinitely strong interactions.

\begin{acknowledgments}
K.\,V.\,K. acknowledges support by the Australian Research Council Discovery Project Grant No. DP190101515.
 \end{acknowledgments}

\appendix

\section{Current density for an ideal Bose gas}
\label{appendix:Probability current for an ideal Bose gas}
Given the time-dependent wave function, Eq.~\eqref{eq:initial_wf}, for the ideal Bose gas, which is of the form 
$\Psi(x,t)\!=\!\psi_{\mathrm{bg}}[1\!+\!B(x,t)e^{i\varphi(x,t)}]$, the current density
\begin{align}
    j(x,t) = \rho(x,t) v(x,t) = \rho(x,t) \frac{\hbar}{m} \frac{\partial S(x,t)}{\partial x}
\end{align}
can be determined using the phase
\begin{align}
	S(x,t) &= \frac{1}{2i} \ln(\frac{\Psi(x,t)}{\Psi^{*}(x,t)}) \nonumber\\
	&= \frac{1}{2i} \ln(\frac{1 + B(x,t)e^{i\varphi(x,t)}}{1 + B(x,t)e^{-i\varphi(x,t)}}).
\end{align}
The velocity is thus given by
\begin{align}
	v(x,t) &= \frac{\hbar}{m} \frac{\partial S(x,t)}{\partial x} \nonumber\\
	&= \frac{\hbar}{m} \left[ \frac{ B'(x,t)  \sin \varphi(x,t)}{1 + 2B(x,t)\cos\varphi(x,t) + B^{2}(x,t)}  \right.\nonumber\\
	&\qquad+\left. \frac{\varphi'(x,t) \left[B(x,t) \cos\varphi(x,t) + B^{2}(x,t) \right]}{1 + 2B(x,t)\cos\varphi(x,t) + B^{2}(x,t)}  \right],
\end{align}
leading to Eq. \eqref{eq:ideal_current}  
\begin{align}
	j(x,t) &= \rho(x,t) v(x,t) \nonumber\\
	&= \frac{\hbar N_{\mathrm{bg}}}{m L} \left\{B'(x,t)\sin\varphi(x,t) \right.\nonumber\\
	&\qquad\quad + \left.\varphi'(x,t)\left[ B(x,t)\cos\varphi(x,t) + B^{2}(x,t)\right]\right\}
\end{align}
with $B'(x,t)$ and $\varphi'(x,t)$ defined in Eqs. \eqref{eq:B'} and \eqref{eq:varphi'}.

\section{Critical height for a vacuum point to occur in the ideal Bose gas}
\label{appendix:Critical height for a vacuum point to occur}

Here we wish to identify, for a given bump width $\sigma$, the necessary height $\beta_{\mathrm{cr}}$ required for a vacuum point to occur in the mean-field GPE approximation, i.e., for the minimum of an interference fringe in the particle number density to touch zero at a given time $t_{\mathrm{vac}}$. The location of each fringe minimum is difficult to find generally, however, we can determine the location of a fringe if it were to touch zero density by considering the function which envelopes the density from below. This is given by
\begin{align}
    \rho_{\mathrm{env_{-}}}(x,t) &= \frac{N_{\mathrm{bg}}}{L}\left[1 - 2B(x,t) + B^{2}(x,t)\right] \nonumber\\
    &= \frac{N_{\mathrm{bg}}}{L}\left[1 - B(x,t)\right]^{2}, \label{eq:rho_env}
\end{align}
where one simply sets $\cos \varphi(x,t)\!=\!-1$ in the actual density [Eq. \eqref{eq:ideal_rho_interference}  ],
\begin{align}
    \rho(x,t) &= \frac{N_{\mathrm{bg}}}{L}\left[1 + 2B(x,t)\cos \varphi(x,t) + B^{2}(x,t)\right].
\end{align}

In general, the density oscillations do not necessarily touch the enveloping function $\rho_{\mathrm{env_{-}}}(x,t)$ at exactly the location of the fringe minimums. This means that the fringe minimums do not correspond exactly with the minimum locations of $\cos \varphi(x,t)$. However, when a vacuum point occurs, the density at that location and time is exactly $\rho(x_{\mathrm{vac}},t_{\mathrm{vac}})\!=\!\rho_{\mathrm{env_{-}}}(x_{\mathrm{vac}},t_{\mathrm{vac}})=0$. Hence, at the vacuum point itself, it must be true that the fringe minimum occurs at exactly the corresponding minimum of $\cos \varphi(x,t)$. This allows one to express the location of a vacuum point using the following steps,
\begin{align}
    -1 &= \cos \varphi(x_{\mathrm{vac}},t_{\mathrm{vac}}),\nonumber\\
    \pi (2n-1) &= \varphi(x_{\mathrm{vac}},t_{\mathrm{vac}}), \nonumber\\
	\pi (2n-1) &= \frac{\hbar t_{\mathrm{vac}} x_{\mathrm{vac}}^{2}}{2m \left(\sigma^{4} + \hbar^{2} t_{\mathrm{vac}}^{2}/m^{2}\right)} - \frac{1}{2} \atan(\frac{\hbar^{2} t_{\mathrm{vac}}^{2}}{m^{2} \sigma^{4}}), \nonumber\\
	x_{\mathrm{vac}}^{2} &= \left[\pi (2n-1) + \frac{1}{2} \atan(\frac{\hbar^{2} t_{\mathrm{vac}}^{2}}{m^{2} \sigma^{4}})\right] \nonumber\\
	&\qquad\qquad\qquad\times\frac{2m \left(\sigma^{4} + \hbar^{2} t_{\mathrm{vac}}^{2}/m^{2}\right)}{\hbar t_{\mathrm{vac}}}, \label{eq:x_vac}
\end{align}
where $n=1,2,3,...$ denotes the fringe number (increasing from the trailing to leading edge) of the shock wave.

Additionally, the requirement $\rho(x_{\mathrm{vac}},t_{\mathrm{vac}})\!=\!\rho_{\mathrm{env_{-}}}(x_{\mathrm{vac}},t_{\mathrm{vac}})=0$ implies that one has
\begin{align}
    \rho_{\mathrm{env_{-}}}(x_{\mathrm{vac}},t_{\mathrm{vac}}) &=\frac{N_{\mathrm{bg}}}{L}\left[1 - B(x_{\mathrm{vac}},t_{\mathrm{vac}})\right]^{2} = 0
\end{align}
at the location of the vacuum point [according to Eq. \eqref{eq:rho_env}] and therefore,
\begin{align}
    B(x_{\mathrm{vac}},t_{\mathrm{vac}}) &= 1.
\end{align}
This then provides a means of determining the value $\beta$ necessary to generate a vacuum point. One uses the requirement that
\begin{align}
	1 &= B(x_{\mathrm{vac}},t_{\mathrm{vac}}) \nonumber\\
	&= \frac{\beta \sigma}{\left(\sigma^{4} + \hbar^{2} t_{\mathrm{vac}}^{2}/m^{2}\right)^{1/4}} e^{-x_{\mathrm{vac}}^{2} \sigma^{2}/2\left(\sigma^{4} + \hbar^{2} t_{\mathrm{vac}}^{2}/m^{2}\right)},
\end{align}
and substitutes in the vacuum point location \eqref{eq:x_vac}. Rearranging the resulting expression then allows one to find the critical value of $\beta$ required for a vacuum point to occur at fringe $n$ at time $t_{\mathrm{vac}}$;
\begin{align}
	\beta_{\mathrm{cr}}^{(n)} &= \frac{1}{\sigma} \left(\sigma^{4} + \hbar^{2} t_{\mathrm{vac}}^{2}/m^{2}\right)^{1/4} \nonumber\\
	&\qquad\qquad\times e^{\frac{m \sigma^{2}}{\hbar t_{\mathrm{vac}}} \left[\pi\left(2n-1\right) + \frac{1}{2}\arctan(\frac{\hbar^{2} t_{\mathrm{vac}}^{2}}{m^{2} \sigma^{4}})\right]}.
\end{align}
This is the formula given in the main text as Eq.~\eqref{eq:beta_cr}, .

\section{The truncated Wigner approach}
\label{appendix:The truncated Wigner approach_currents}
For the results presented in this work that incorporate quantum fluctuations, the truncated Wigner approach (TWA) \cite{Steel1998,Blakie2008,Martin2010}
is used with Bogoliubov initial conditions. Specifically, the 1D version of the Bogoliubov--de Gennes (BdG) equations for the mode amplitudes $u_{j}(x,0)$ and $v_{j}(x,0)$ \cite{Pethick&Smith2008,Pitaevskii&Stringari2016,Stringari1999,Hu2015} 
\begin{align}
    \mu\Psi_{0}
	&= \left[ \frac{-\hbar^{2}}{2m}\frac{\partial^{2}}{\partial x^{2}} + V + g_{\mathrm{1D}} \left| \Psi_{0} \right|^{2} \right] \Psi_{0}, \label{eq:ti-GPE}\\
	E_{j}u_{j} &= \left(\frac{-\hbar^{2}}{2m} \frac{\partial^{2}}{\partial x^{2}} + V -\mu + 2g_{\mathrm{1D}}|\Psi_{0}|^{2}\right) u_{j}\nonumber\\
	&\qquad\qquad\qquad\qquad\qquad\qquad+ g_{\mathrm{1D}}\Psi_{0}^{2} v_{j},\label{eq:BdG_u}\\
	-E_{j}v_{j} &= \left(\frac{-\hbar^{2}}{2m} \frac{\partial^{2}}{\partial x^{2}} + V -\mu + 2g_{\mathrm{1D}}|\Psi_{0}|^{2}\right) v_{j}\nonumber\\
	&\qquad\qquad\qquad\qquad\qquad\qquad+ g_{\mathrm{1D}}(\Psi_{0}^{*})^{2} u_{j}, \label{eq:BdG_v}
\end{align}
(where $\mu$ is the chemical potential of the condensate mode $\Psi_{0}$, and $E_{j}$ the energy of the $j^{\text{th}}$ Bogoliubov mode) are solved simultaneously using the trapping potential \cite{Simmons2020}
\begin{align}
    V(x) &= \frac{\hbar^{2}\beta}{2m\sigma^{4}} \left(x^{2} - \sigma^{2}\right) \left[e^{x^{2}/2\sigma^{2}} + \beta\right]^{-1} \nonumber\\
    &\qquad\qquad\qquad- \frac{g_{\mathrm{1D}}N_{\text{bg}}}{L}\left[1+\beta e^{-x^{2}/2\sigma^{2}}\right]^{2} \label{eq:V_trap-appendix_Currents}
\end{align}
in order to obtain the initial state of the system, which includes both the condensate ($j=0$) and non-condensate ($j=1,2,3,...$) modes.

We note that the BdG equations admit both positive and negative energy solutions. Specifically, for each solution with positive energy, $E_{j}$, there is an equivalent solution with negative energy, $-E_{j}$, that provides an equivalent physical description. In practice then, one need only work with, say, the positive solutions. Since the number of particles in the non-condensed Bogoliubov field is given by $N_{\mathrm{nc}}=\int \langle\delta\hat{\psi}^{\dagger}(x,0)\delta\hat{\psi}(x,0)\rangle\,dx = \int \sum_{j\geq1}|v_{j}(x,0)|^{2}\,dx$, the total number of particles in the system can be computed using $N=N_{0}+N_{\mathrm{nc}}=\int (|\Psi_{0}(x,0)|^{2} + \sum_{j\geq1}|v_{j}(x,0)|^{2})\,dx$. In practice, the sum over excited states used to compute $N_{\mathrm{nc}}$ must be truncated to some finite number of modes $M$.

The trapping potential \eqref{eq:V_trap-appendix_Currents} was originally derived in Appendix II of Ref. \cite{Simmons2020}, but we now allow $\beta<0$ in order to generate initial density dips as well. This trap produces exactly the desired density $\rho_{0}(x,0)=N_{\text{bg}}[1+\beta e^{-x^{2}/2\sigma^{2}}]^{2}/L$ as the ground state solution to the GPE \eqref{eq:ti-GPE}. Therefore, the population of the excited Bogoliubov modes does cause small deviations from this desired initial density. Provided the total number of non-condensed atoms $N_{\mathrm{nc}}$ remains sufficiently small compared with the total number of condensed atoms $N_{0}$, this deviation should be negligible. This is the case for the scenarios we examine in this work, i.e., $N_{\mathrm{nc}}\ll N_{0}$.

Following diagonalization of the BdG equations, each stochastic trajectory is then initialized according to the Bogoliubov prescription \cite{Steel1998,Blakie2008,Martin2010}
\begin{equation}
  \psi_{W}(x,0) = \Psi_0(x,0) + \sum_{j=1}^{M} \left[\eta_{j} u_{j}(x) + \eta_{j}^{*}v^*_{j}(x)\right],
\end{equation}
and evolved according to the time-dependent GPE \eqref{eq:td-GPE}. Here, $\eta_{j}$ is a complex Gaussian noise term with zero mean, $\langle \eta_{j}\rangle_{\mathrm{stoch}}= 0$, and variance $\langle\eta_{j}^{*}\eta_{k}\rangle_{\mathrm{stoch}} = \frac{1}{2}\delta_{jk}$. Moreover, $\Psi_0(x,0)$ is the ground state of the time-independent GPE \eqref{eq:ti-GPE} which we renormalize to the occupation of the condensate mode $N_{0}$ so that the total number of atoms $N=N_0+N_{\mathrm{nc}}$ remains the same as in the mean-field GPE simulations with which we compare the TWA results to in the main text.

We note that, for the simulation presented in Figs. \ref{fig:Currents_N2000_VacuumPoint2}\,(a) and (d) the mode population used in the TWA was $N/M\approx5$. Similarly: for Figs. \ref{fig:Currents_N2000_VacuumPoint2}\,(b) and (e), $N/M\approx1$; for Figs. \ref{fig:Currents_N2000_sigma_sweep}\,(a) and (d), $N/M\approx3.4$; for Figs. \ref{fig:Currents_N2000_sigma_sweep}\,(b) and (e), $N/M\approx3.3$; and  for Figs. \ref{fig:Currents_N2000_sigma_sweep}\,(c) and (f), $N/M\approx3.2$. While some of these may seem rather small, it has been made known in the literature that the effect of varying the mode occupation for TWA in one-dimension has a very limited effect (if any at all) on the results \cite{Ruostekoski2005,Ruostekoski2006,Ruostekoski2012,Simmons2022}.

Furthermore, in the truncated Wigner approach, stochastic averages over phase-space variables correspond to symmetrically ordered expectation values of their respective operators such that one would have, for example,
\begin{align}
    \langle\hat{\Psi}^{\dagger}(x,t)\hat{\Psi}(x,t)\rangle_{\mathrm{sym}}
    &= \frac{1}{2}\langle\hat{\Psi}^{\dagger}(x,t) \hat{\Psi}(x,t) \nonumber\\
    &\qquad\qquad\qquad+ \hat{\Psi}(x,t) \hat{\Psi}^{\dagger}(x,t)\rangle\nonumber\\
    &= \langle\widetilde{\psi}_{W}(x,t)\psi_{W}(x,t)\rangle.\label{eq:TWA c-field correspondence}
\end{align}
This means that, to compute observables which are normally ordered, one needs to employ the Bose commutation relations to correctly represent the symmetrically ordered operator expectation values of the Wigner approach. For the density itself, this results in
\begin{align}
    \rho(x,t)&=\langle\hat{\Psi}^{\dagger}(x,t)\hat{\Psi}(x,t)\rangle\\
	&=\langle\psi_{W}^{*}(x,t)\psi_{W}(x,t)\rangle-\frac{1}{2}\delta_{c}(x,x), \label{eq:TWA_density-delta}
\end{align}
where $\frac{1}{2}\int\delta_{c}(x,x)\,dx=\frac{1}{2}M$ represents the half quantum of vacuum noise per mode $M$ included in the Wigner formalism, which must now be subtracted off to obtain the mean density of the gas.

Using an appropriate computational grid, which has spacing $\Delta x = L/\mathcal{N}_{x}$, with $\mathcal{N}_{x}$ being the number of grid points, the projected delta function $\delta_{c}(x,x)$ of Eq. \eqref{eq:TWA_density-delta} is simply $\delta_{c}(x,x)=M/(\mathcal{N}_{x}\Delta x)=M/L$ \cite{Blakie2008}. In this case one can compute the density in the TWA approach using
\begin{align}
	\rho(x,t) & \equiv\langle\hat{\Psi}^{\dagger}(x,t)\hat{\Psi}(x,t)\rangle\nonumber\\
	&=\langle|\psi_{W}(x,t)|^{2}\rangle-\frac{M}{2L}.
\end{align}
On the other hand, the current density can be computed using \cite{Drummond2017},
\begin{align}
    j(x,t) &= \frac{\hbar}{2mi} \left<\hat{\Psi}^{\dagger}(x,t)\frac{\partial \hat{\Psi}(x,t)}{\partial x} - \hat{\Psi}(x,t)\frac{\partial \hat{\Psi}^{\dagger}(x,t)}{\partial x}\right>\\
    &= \frac{\hbar}{2mi} \left<\psi_{W}^{*}(x,t)\frac{\partial}{\partial x} \psi_{W}(x,t)\right. \nonumber\\
    &\qquad\qquad\qquad\qquad- \left.\psi_{W}(x,t)\frac{\partial}{\partial x} \psi_{W}^{*}(x,t)\right>\\
    &= \frac{\hbar}{m} \Im \left<\psi_{W}^{*}(x,t)\frac{\partial}{\partial x} \psi_{W}(x,t)\right>. \label{eq:j_TWA}
\end{align}

\section{The stochastic projected Gross-Pitaevskii equation (SPGPE)}\label{appendix:SPGPE}

Simulations of shock wave dynamics initiated from a thermal equilibrium state at a nonzero initial temperature are performed using the stochastic Gross-Pitaevskii approach \cite{Castin2000,Davis2001,Blakie2008}; here we follow the prescription detailed in Appendix III of Ref. \cite{Simmons2020}. The essence of this approach is to treat the highly occupied modes of the Bose field operator as a classical field. The initial thermal equilibrium state of this classical field is determined using the simple growth stochastic projected Gross-Pitaevskii equation (SPGPE) \cite{Blakie2008} in which the field is coupled to a thermal reservoir (the low-occupancy modes) at temperature $T$. This initial thermal equilibrium state is prepared in the trapping potential given by Eq. \eqref{eq:V_trap-appendix_Currents} and the stochastic field realizations are then evolved in time according to the standard GPE \eqref{eq:td-GPE}, with $V(x)=0$, in order to determine the actual shock wave dynamics. The current density can be computed according to Eq. \eqref{eq:j_TWA} above where the Wigner field $\psi_{W}(x,t)$ is replaced by the classical field of the SPGPE prescription.

We note that the SPGPE is valid for degenerate 1D Bose gases with temperatures in the range $\gamma\lesssim \overline{\mathcal{T}} \lesssim 1$, where $\overline{\mathcal{T}}=T/T_{d}$ and $T_{d}=\hbar^{2}\rho^{2}/2mk_{B}$ is the temperature of quantum degeneracy ($k_{B}$ is Boltzmann's constant). For non-uniform gases like we consider here, one should ensure that the temperature $\overline{\mathcal{T}}(x)$ remains within these bounds across the entirety (or at least the majority) of the system, since $\rho$ and therefore $\gamma$ vary spatially across the gas. One way to do this is to consider the spatially invariant temperature $\overline{\tau}=k_{B}T/(mg_{\mathrm{1D}}^{2}/2\hbar^{2})=\overline{\mathcal{T}}/\gamma^{2}$ and choose temperatures according to the bounds $\gamma^{-1}\lesssim \overline{\tau} \lesssim \gamma^{-2}$. This is what we have done in this work. For further discussion on these regimes of validity we refer the reader to Appendix A of Ref. \cite{Bayocboc2022}.

\section{Exact diagonalization in the TG limit}
\label{appendix:Exact diagonalisation in the TG limit}

In the TG limit we employ the Fermi-Bose mapping of Refs. \cite{Girardeau1960,Yukalov2005} that allows us to describe the many-body dynamics of the TG gas in terms of single-particle wave functions of a free-fermion system. In particular, the $N$-body wave function $\Psi$ of the TG gas can be obtained via
\begin{equation}
  \Psi(x_{1},\ldots,x_{N};t) = A(x_{1},\ldots,x_{N}) \Psi^{(F)}(x_{1},\ldots,x_{N};t),
\end{equation}
where $\Psi^{(F)}$ represents a spin-less free fermion $N$-body wave function for the same dynamical scenario, and the unit antisymmetric function is given by $A(x_{1},\ldots,x_{N}) = \prod_{1\leq j < i \leq N} \mathrm{sgn}(x_{i} - x_{j})$ which ensures the correct symmetrization of the bosonic wave function. Moreover, the $N$-body fermionic wave function itself can be constructed using the Slater determinant
\begin{equation}
  \Psi^{(F)}(x_{1},\ldots,x_{N};t) = \frac{1}{\sqrt{(N-1)!}} \det_{i,j=1}^{N}[\phi_{i}(x_{j},t)], \label{eq:Slater_det-appendix_currents}
\end{equation} 
where the single-particle wave functions $\phi_{i}(x,t)$ evolve according to the Schr\"odinger equation from their initial states $\phi_{i}(x,0)$, which are eigenstates of the initial trapping potential $V(x,0)$ with eigenenergies $E_{i}$ such that the total energy of the $N$-body wave function is $E_{\mathrm{TG}} = \sum_{i=1}^{N} E_{i}$. For further details regarding the construction of the TG solution, as well as the required initial trapping potential $V(x,0)$ for the shock wave scenarios we consider, see Refs. \cite{Girardeau1960,Girardeau2002,Yukalov2005,Simmons2020}.

While it is known that the mean particle number density of the TG gas can be computed using
\begin{align}
  \rho(x,t) &= \sum_{i=1}^{N}
  \left|\phi_{i}(x,t)\right|^{2}, \label{eq:TG_density-appendix_currents}
\end{align}
here we provide a derivation of how one can construct the mean probability current $j(x,t)$ from free-fermion wave functions.

We begin by first writing this quantity in terms of the Slater determinant of single-particle wave functions,
\begin{align}
    j(x_{1};t)     &= \frac{\hbar}{m} \Im \left( \int \Psi^{*}(x_{1},x_{2},...,x_{N};t)\right. \nonumber\\
    &\qquad\qquad\qquad\times\left.\frac{\partial \Psi(x_{1},x_{2},...,x_{N};t)}{\partial x_{1}} dx_{2}...dx_{N}\right) \nonumber\\
    	&= \frac{\hbar}{m} \Im \left( \int \Psi^{*(F)}(x_{1},x_{2},...,x_{N};t) \right. \nonumber\\
	&\qquad\qquad\quad\times \left.\frac{\partial \Psi^{(F)}(x_{1},x_{2},...,x_{N};t)}{\partial x_{1}} dx_{2}...dx_{N}\right) \nonumber\\
	&= \frac{\hbar}{m} \Im \left( \frac{1}{(N-1)!}\int \det_{i,j=1}^{N}[\phi^{*}_{i}(x_{j};t)] \right. \nonumber\\
	&\qquad\qquad\quad\times \left.\frac{\partial \det_{k,l=1}^{N}[\phi_{k}(x_{l};t)]}{\partial x_{1}} dx_{2}...dx_{N}\right), \label{eq:TG_current_det}
\end{align}
where we have used $A^{2}(x_{1},x_{2},...,x_{N})=1$. At this point it is possible to significantly simply the result by considering the terms which will arise in the integrand. 

Since any single-particle wave function with $\phi_{k}(x_{l\neq1};t)$ can be pulled out of the derivative $\partial/\partial x_{1}$, we note that the integral over $x_{2}...x_{N}$ will kill any cross terms for which the index pairs $(k,l)$ don't match $(i,j)$, due to the orthonormality of the single-particle wave functions. Moreover, this orthonormality will leave a summation of many other terms, each of which can be written as $\phi^{*}_{i}(x_{1};t) \frac{\partial}{\partial x_{1}}\phi_{i}(x_{1};t)$, where the pre-factors of these remaining terms will integrate to one. In fact, there will be precisely $(N-1)!$ of these terms for each $i$. That is, there will be $(N-1)!$ of the terms above for each of the $N$ single-particle wave functions, and this leads to
\begin{align}
    \int& \det_{i,j=1}^{N}[\phi^{*}_{i}(x_{j};t)] \frac{\partial \det_{k,l=1}^{N}[\phi_{k}(x_{l};t)]}{\partial x_{1}} dx_{2}...dx_{N} \nonumber\\
    &\qquad\qquad= (N-1)! \sum_{i=1}^{N} \phi^{*}_{i}(x_{1};t) \frac{\partial}{\partial x_{1}}\phi_{i}(x_{1};t). \label{eq:TG_current_remainder}
\end{align}

Substituting Eq.~\eqref{eq:TG_current_remainder} back into Eq.~\eqref{eq:TG_current_det} shows that it is possible to compute the probability current of a zero-temperature TG gas using
\begin{align}
    j(x,t) &= \frac{\hbar}{m} \Im \left( \sum_{i=1}^{N} \phi_{i}^{*}(x,t) \frac{\partial}{\partial x} \phi_{i}(x,t)\right), \label{eq:j_TG_Background-appendix_currents}
\end{align}
which is nothing more than a summation over the single-particle currents \cite{Dubessy2021,Minguzzi2022},
\begin{align}
    j_{i}(x,t)&= \frac{\hbar}{m} \Im \left( \phi_{i}^{*}(x,t) \frac{\partial}{\partial x} \phi_{i}(x,t)\right).
\end{align}

In order to actually compute the particle number density and current density, we use the same methodology as in Ref. \cite{Simmons2020}. Specifically, we obtain the initial and time-evolved wave functions $\phi_{i}(x,t)$ using the trapping potentials and procedures outlined in the Supplemental Material of that work, where we use $\beta>0$ to generate traps which create initial density bumps, and $\beta<0$ to generate traps which create initial density dips. We use Eq. (S33) of Ref. \cite{Simmons2020} for scenarios with $N\geq40$ and Eq. (S9) [with $\bar{g}=0$] for scenarios with $N=4$.

\begin{widetext}
\begin{figure*}[tbp]
    \centering
    \includegraphics[width=0.99\linewidth]{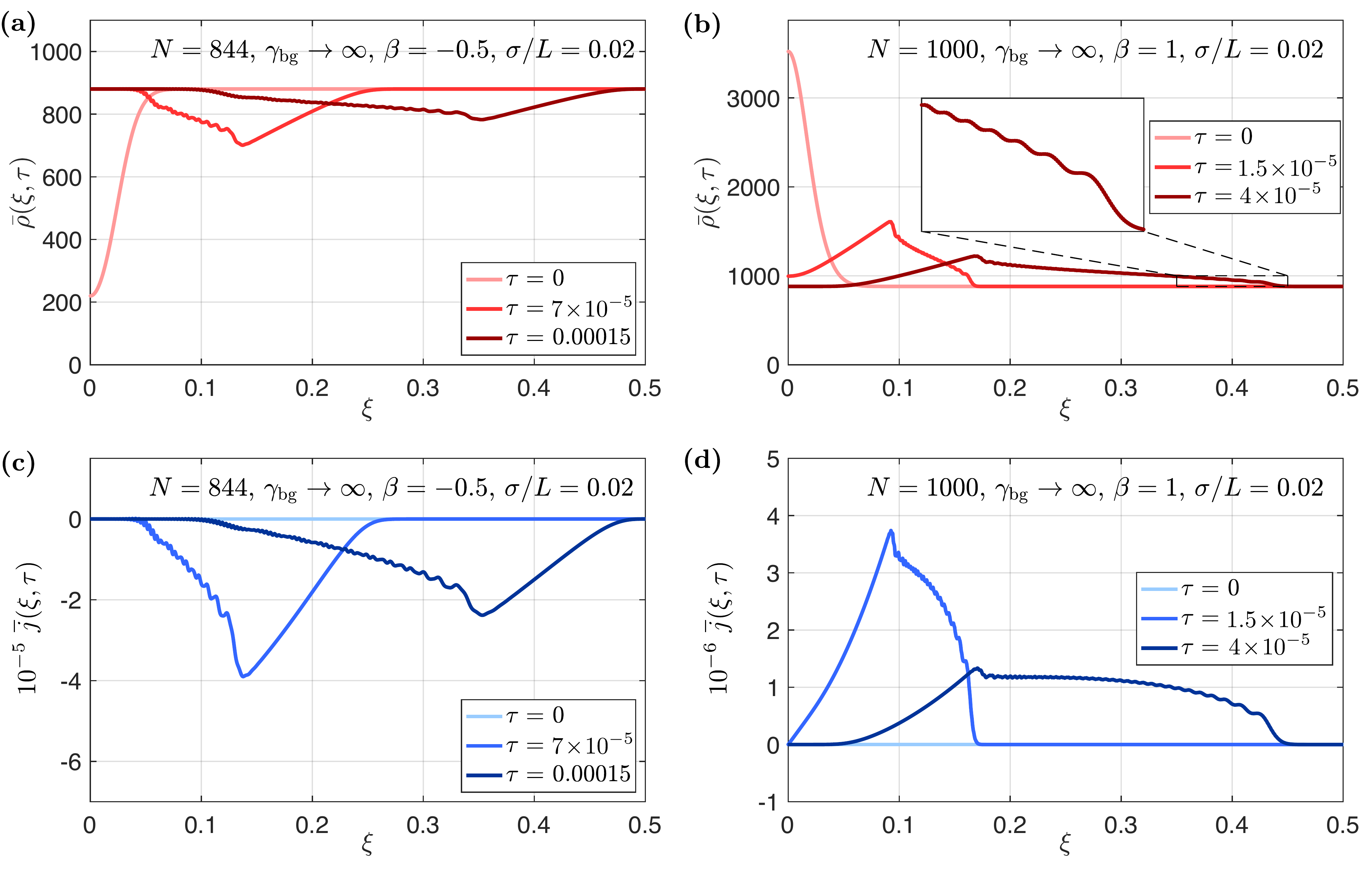}
    \caption{Shock waves generated in the TG regime of a zero-temperature 1D Bose gas. The probability densities (top row) and currents (bottom row) are shown for: (a) and (c)---a density dip scenario with $N=844$, $\gamma_{\mathrm{bg}}\rightarrow\infty$, $\beta=-0.5$, $\sigma/L=0.02$, leading to $N_{\mathrm{bg}}=880.3$; (b) and (d)---a density bump scenario with $N=1000$, $\gamma_{\mathrm{bg}}\rightarrow\infty$, $\beta=1$, $\sigma/L=0.02$, leading to $N_{\mathrm{bg}}=880.5$. In (a) identifiable longer-wavelength Bettelheim-Glazman oscillations appear on the left edge of the shock, whereas the inset in (b) shows more clearly the appearance of many such oscillations (that span numerous Friedel oscillations) in the scenario plotted there.
    }
    \label{fig:Currents_N1000_TG_limit}
\end{figure*}
\end{widetext}

\section{Bettelheim-Glazman oscillations in the TG limit}
\label{appendix:Bettelheim-Glazman oscillations in the TG limit}

In the TG limit, for large particle numbers $N$, Bettelheim-Glazman oscillations \cite{Bettelheim2012,Glazman2019} are present in the shock wave envelope and we identify them in this Appendix. In Figs. \ref{fig:Currents_N1000_TG_limit}\,(a) and (c) we show the same scenario as in Fig. \ref{fig:Currents_N50Dip_Interaction_Sweep}\,(c)   (which was for $N=42$) but now for $N=844$ particles and with $N_{\mathrm{bg}}\simeq880$. In this density dip scenario, we identify Bettelheim-Glazman oscillations as density modulations in the shock wave envelope whose wavelength becomes larger towards both the leading (right) and trailing (left) edges of the envelope. That is to say, the frequency of oscillations is chirped toward the middle of the shock. This is in contrast to the chirp direction of the interference pattern seen in the Thomas-Fermi regime of a weakly interacting system, which is towards the leading edge of the shock wave.

In an equivalent density bump scenario, with $N_{\mathrm{bg}}\simeq880$ but now $N=1000$, shown in Figs. \ref{fig:Currents_N1000_TG_limit}\,(b) and (d), we see the same Bettelheim-Glazman oscillations: their frequency is chirped in the same way as before, i.e., from the trailing and leading edges of the shock wave envelope towards the middle. The inset of Fig. \ref{fig:Currents_N1000_TG_limit}\,(b) shows these oscillations more clearly in the leading  edge of the shock wave envelope. They span many of the barely visible Friedel oscillations whose amplitude is on the order of the linewidth, and whose period $\sim1/\rho_{\mathrm{bg}}$ is also not visible on the scale of the figure.

Bettelheim-Glazman oscillations were first predicted in Ref. \cite{Bettelheim2012} by considering leading order quantum corrections in the Wigner function representation of an ideal Fermi gas, whose particle number density can be mapped to the density of a bosonic TG gas. The results presented here confirm the existence of these oscillations in an exact quantum many-body calculation. By going to a larger total number of particles compared with the cases considered in Ref.~\cite{Simmons2020} we have thus been able to discern between Bettelheim-Glazman and Friedel-type oscillations, as the characteristic period and amplitude of the former are now much larger than those of Friedel oscillations.

We should point out that our simulations are not strictly in the regime considered in Ref. \cite{Bettelheim2012,Glazman2019}; where the density bump should be small compared to the background density, yet still contain many particles itself. As such, our results possess some of the Friedel-type oscillations between different branches of the Fermi momentum (in phase-space) which are described in Ref. \cite{Protopopov2013}. These oscillations cause the irregular peaks which are visible across the shock front. Nevertheless, we still see the clear signature of longer-wavelength Bettelheim-Glazman oscillations appearing on top of these features.

\section{Identification of solitons in the mean-field GPE}
\label{appendix:Identification of solitons in the mean-field GPE}
Here we explain how we identified rigorously the presence of solitons in the results of Figure \ref{fig:Currents_N2000_sigma_sweep}.

In the variables used throughout this work, the analytic profile of a gray soliton whose minimum is initially ($t=0$) positioned at the origin and then proceeds to travel to the right can be written as \cite{Pitaevskii&Stringari2016}
\begin{align}
    \Psi_{\mathrm{s}} &= \sqrt{\frac{N_{\mathrm{bg}}}{L}} \left(i\frac{v}{c_{\mathrm{bg}}} + \sqrt{1-
    \frac{v^{2}}{c_{\mathrm{bg}}^{2}}}\right. \nonumber\\
	&\qquad\qquad\qquad\quad\times \left. \tanh\left[\frac{x-vt}{l_{\mathrm{h}}} \sqrt{1-
    \frac{v^{2}}{c_{\mathrm{bg}}^{2}}}\right]\right), \label{eq:psi_soliton}
\end{align}
where we have the background particle number $N_{\mathrm{bg}}$, speed of sound in the background $c_{\mathrm{bg}}=\sqrt{g_{\mathrm{1D}}\rho_{\mathrm{bg}}/m}=\hbar N_{\mathrm{bg}}\sqrt{\gamma_{\mathrm{bg}}}/mL$, and the healing length in the background $l_{\mathrm{h}}=\hbar/\sqrt{mg_{\mathrm{1D}}\rho_{\mathrm{bg}}}=L/N_{\mathrm{bg}}\sqrt{\gamma_{\mathrm{bg}}}$. Additionally, the velocity with which the soliton propagates (undisturbed) through the fluid can be computed using \cite{Pitaevskii&Stringari2016}
\begin{align}
    v^{2} &= c_{\mathrm{bg}}^{2} \frac{\rho_{\mathrm{min}}}{\rho_\mathrm{bg}} = c_{\mathrm{bg}}^{2} \frac{N_{\mathrm{min}}}{N_\mathrm{bg}},
\end{align}
which requires us to determine the density at the minimum of the soliton.

\begin{figure}[tbp]
    \centering
    \includegraphics[width=\linewidth]{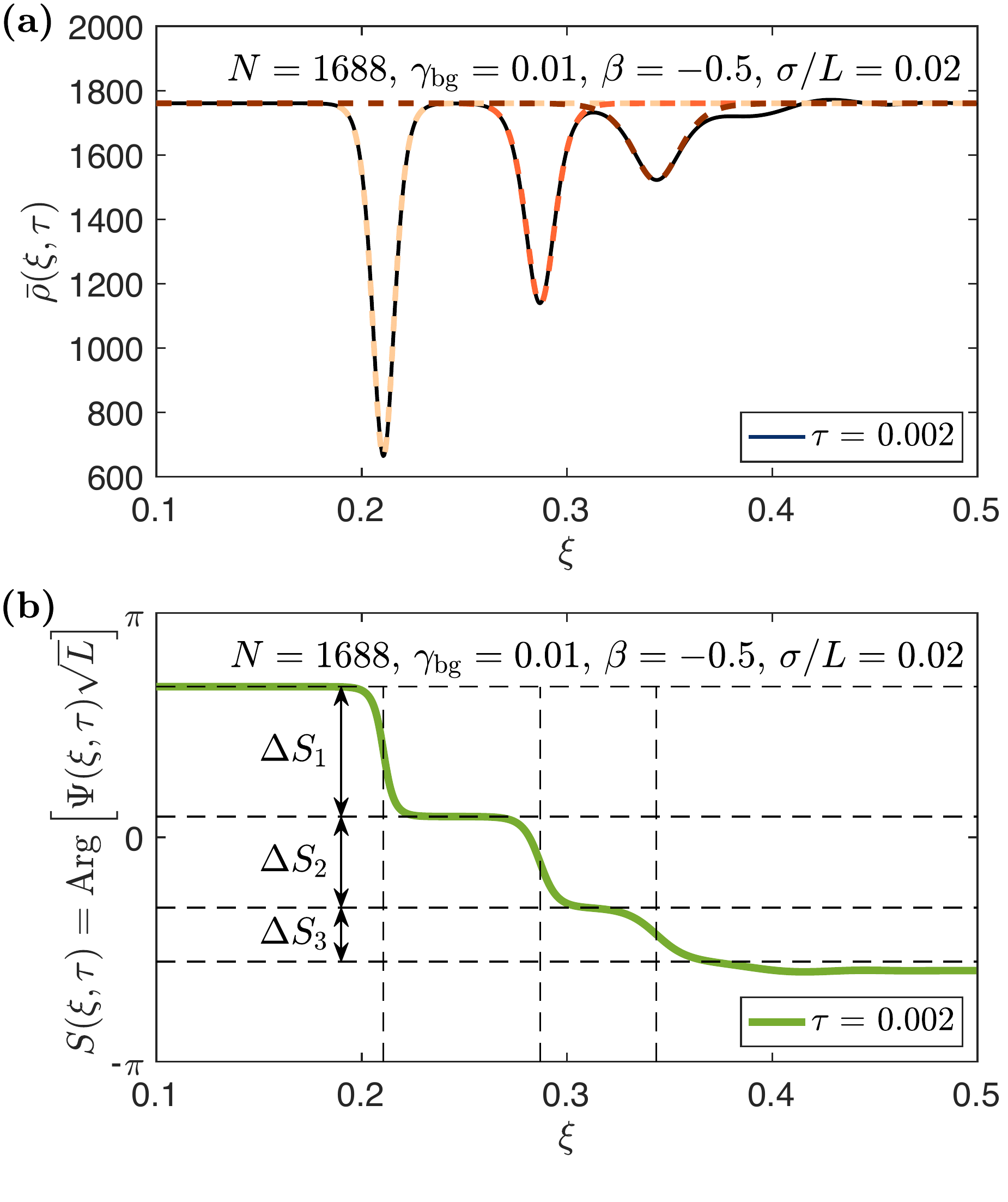}
    \caption{Mean-field GPE density and phase profiles at time $\tau=0.002$ for the dynamical situation of Fig. \ref{fig:Currents_N2000_sigma_sweep}\,(b). Panel (a) provides the dimensionless density where the dashed lines correspond to fits of soliton profiles for each of the density dips (shifted so that the minimums match). The soliton fits match the density profile extremely well, indicating that it is correct to identify these structures as solitons. Panel (b) shows the phase profile where the vertical dashed lines correspond to the position of density minimums of panel (a), and the horizontal dashed lines denote the predicted soliton phase shifts across each dip. Again, the predicted phase shifts match very well with the phase drops across each dip, further confirming that these structures are solitons.
    }
    \label{fig:Currents_Soliton_Density_Phase}
\end{figure}

Hence, in order to determine the presence of solitons we first determine the minimum density of any solitons that are shed from the initial density dip. Along with the knowledge of $N_{\mathrm{bg}}$ and $\gamma_{\mathrm{bg}}$ this then allows us to calculate the solitons velocity and construct the density profile \eqref{eq:psi_soliton}, that we manually shift to the location of the respective soliton minimum, $x_{0}$, using $x-vt\rightarrow x-x_{0}$. As an example we choose the situation of Fig. \ref{fig:Currents_N2000_sigma_sweep}\,(b) and show the outcome of such a fitting procedure in Fig. \ref{fig:Currents_Soliton_Density_Phase}\,(a)---for all three density dips in the soliton train that can individually be identified as a gray soliton. As we see, the agreement is excellent, even for the solitons which have not yet completely separated from each other.

In addition to the profile of Eq. \eqref{eq:psi_soliton}, solitons also have a well-defined phase shift across their profile \cite{Pitaevskii&Stringari2016},
\begin{align}
    \Delta S = 2\arccos\left(\frac{v}{c_{\mathrm{bg}}}\right), \label{eq:phase_soliton}
\end{align}
and we calculate this across each soliton in the example provided here.

From left to right in Fig. \ref{fig:Currents_Soliton_Density_Phase}\,(a) the minimum density of each soliton is $\rho_{\mathrm{min}}^{(1)}L\simeq662.6$, $\rho_{\mathrm{min}}^{(2)}L\simeq1137.7$, $\rho_{\mathrm{min}}^{(3)}L\simeq1520.7$. Using $N_{\mathrm{bg}}\simeq1761$ and $\gamma_{\mathrm{bg}}=0.01$ leads to a dimensionless background speed of sound $c_{\mathrm{bg}}mL/\hbar\simeq176.1$, and therefore the dimensionless velocity ($\bar{v}=vmL/\hbar$) of each soliton is $\bar{v}_{1}\simeq108.0$, $\bar{v}_{2}\simeq141.5$, $\bar{v}_{3}\simeq163.6$. Substituting these into Eq. \eqref{eq:phase_soliton} gives the predicted phase shifts across the solitons: $\Delta S_{1}\simeq1.821$, $\Delta S_{2}\simeq1.274$, and $\Delta S_{3}\simeq0.756$ radians. In Fig. \ref{fig:Currents_Soliton_Density_Phase}\,(b) we plot the actual phase profile of the Bose gas from the mean-field GPE simulations at the same simulation time $\tau=0.002$ as the density in Fig. \ref{fig:Currents_Soliton_Density_Phase}\,(a). Here we determine the `global' phase of the gas from the center of the box ($x=0$), and starting at this value we subtract off consecutively the calculated phase shifts and denote each of those values with horizontal dashed lines. We see that these analytically predicted phase shifts agree extremely well with the actual drops in the phase profile of the gas, further confirming that these structures are solitons.

\begin{figure}[tbp]
    \centering
    \includegraphics[width=1.0\linewidth]{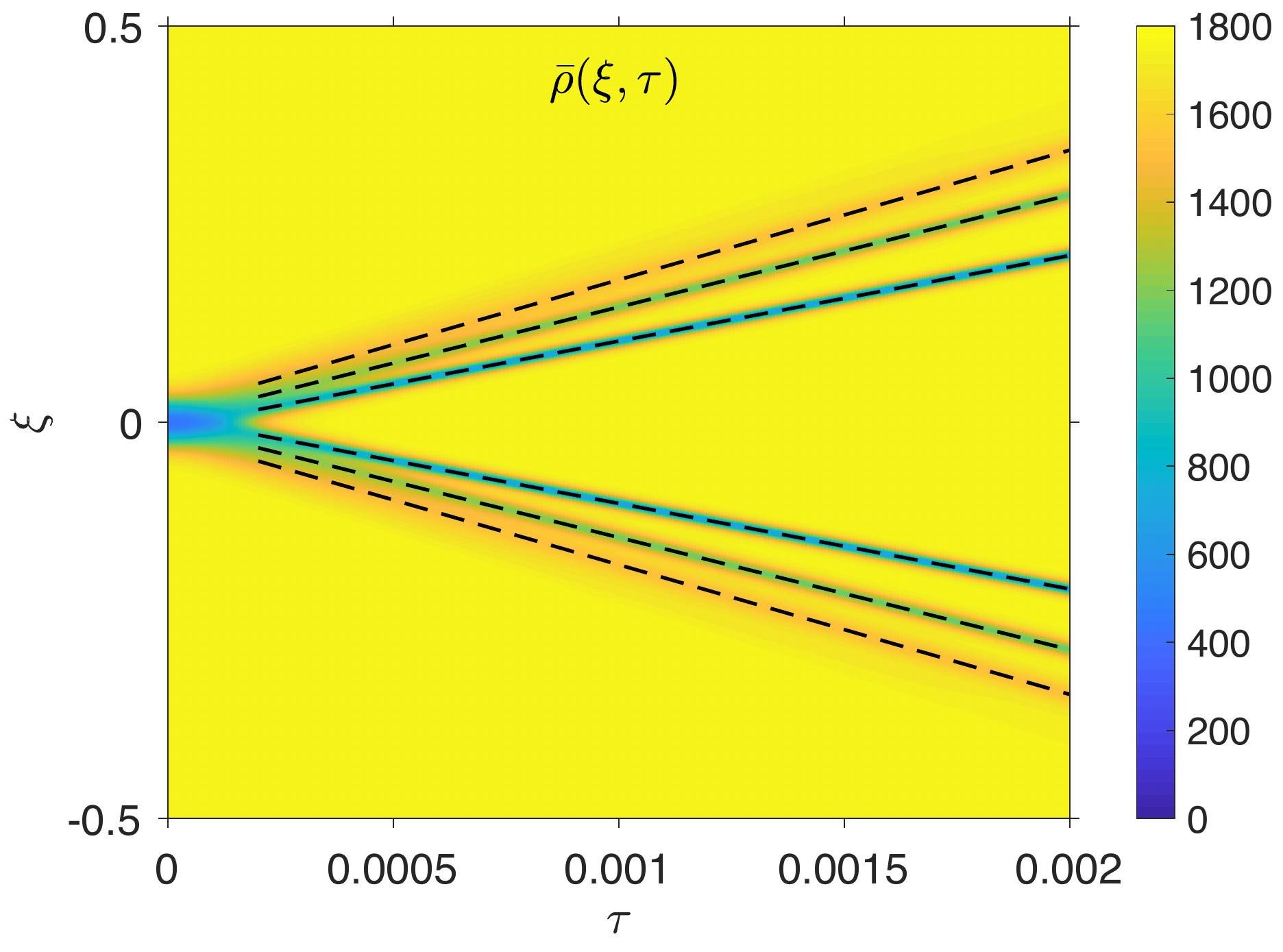}
    \caption{Evolution of the entire density profile for the situation given in Fig. \ref{fig:Currents_N2000_sigma_sweep}\,(b). The slope of the dashed lines are given by the calculated soliton velocities $\bar{v}_{1}\simeq\pm108.0$, $\bar{v}_{2}\simeq\pm141.5$, $\bar{v}_{3}\simeq\pm163.6$, where each line is shifted so as to match with the location of each soliton at the final time $\tau=0.002$.
    }
    \label{fig:Currents_Soliton_Velocity}
\end{figure}

Finally, in Fig. \ref{fig:Currents_Soliton_Velocity} we show that these solitons form and become stable very quickly after being shed from the initial density dip. We plot the density profile over the entire domain $x/L\in[-0.5,0.5]$ as a function of time, which allows us to identify that the density dips travel with a preserved shape and constant velocity almost immediately after the initial density dip separates into left and right moving parts. The slope of the dashed lines indicate the velocities calculated above and they are given by $x=\pm\bar{v}L\tau + x_{0}$, where each is shifted by some different amount $x_{0}$ so that they match with the location of the solitons at time $\tau=0.002$.


\begin{thebibliography}{83}
\bibitem [{cur()}]{current_density_vs_probability_current}%
  \BibitemOpen
  \href@noop {} {}\bibinfo {howpublished} {When the current density
  \eqref{eq:j_0} is normalised to one, it is instead usually referred to as the
  probability current. Moreover, if the number density $\rho_{0}$ used here was
  replaced by the mass density (related simply by a factor of $m$), the current
  density is then often referred to as the mass flux density.}\BibitemShut
  {Stop}%
\bibitem [{sol()}]{soliton_speed}%
  \BibitemOpen
  \href@noop {} {}\bibinfo {howpublished} {The speed of propagation of a
  soliton is related to its depth, or minimum density $\rho_{\mathrm{min}}$,
  through $v_\mathrm{soliton}=\sqrt{g_{\mathrm{1D}}\rho_{\mathrm{min}}/m}$,
  which is exactly the bulk sound velocity evaluated at the density
  $\rho_{\mathrm{min}}$.}\BibitemShut {Stop}%
\end{thebibliography}

%

\end{document}